\documentclass[a4paper,12pt]{article}
\pdfoutput=1 

\usepackage{jheppub} 
                     
\usepackage{bbold}                     
\usepackage{lmodern} 
\usepackage[T1]{fontenc} 
\usepackage[utf8]{inputenc} 

\newcommand{\g}{\gamma}
\newcommand{\ta}{\theta}
\newcommand{\la}{\lambda}
\newcommand{\pd}{\partial}
\newcommand{\al}{\alpha}
\newcommand{\bt}{\beta}

\newcommand{\ep}{\epsilon}
\newcommand{\sg}{\sigma}
\newcommand{\dt}{\delta}

\newcommand{\nn}{\nonumber}

\newcommand{\tah}{\widehat{\theta}}

\newcommand{\dth}{\widehat{\delta}}

\newcommand{\ab}{\underline{a}}
\newcommand{\bb}{\underline{b}}
\newcommand{\cb}{\underline{c}}
\newcommand{\db}{\underline{d}}
\newcommand{\eb}{\underline{e}}
\newcommand{\fb}{\underline{f}}

\newcommand{\pb}{\overline{\partial}}

\newcommand{\zb}{\overline{z}}

\DeclareMathOperator{\sTr}{sTr}
\usepackage[compat=1.1.0]{tikz-feynman}
\usepackage{subcaption}
\usepackage{esvect}

\title{\boldmath Integrating out the fermions in $\rm AdS$}

\author{Cassiano A. Daniel}

\affiliation{ICTP South American Institute for Fundamental Research
\\Instituto de F\'isica Te\'orica, Universidade Estadual Paulista, \\
Rua Dr. Bento Teobaldo Ferraz 271, 01140-070, S\~ao Paulo - SP, Brasil}

\emailAdd{c.daniel@unesp.br}

\abstract{Not much is known about superstring scattering amplitudes in curved backgrounds. Using the hybrid formalism in $\rm AdS_3 \times S^3$ with pure NS-NS three-form flux, we compute a $\rm PSU(1,1|2)$-covariant three-point amplitude for half-BPS vertex operators inserted on the $\rm AdS_3$ boundary and show that it agrees with the RNS computation. The zero-mode prescription for the fermions in $\rm AdS$ is defined in terms of the ``standard'' spacetime SUSY generator. It is found that integrating out the fermionic worldsheet fields in the path integral gives rise to the target-space vielbein, which explicitly encodes that the conformal group on the boundary is identified with the symmetry group of the $\rm AdS$ bulk.}

\begin{document} 
\maketitle
\flushbottom

\section{Introduction}

Establishing a thorough understanding of covariant superstring formulations in Anti de-Sitter (AdS) backgrounds is a much needed step for a deeper comprehension of the dualities connecting the superstring with conformal field theories (CFTs) living on the boundary of the AdS spacetime. Even though a handful of covariant sigma-model actions are explicitly known \cite{Berkovits:1999xv} \cite{Berkovits:1999zq} \cite{Berkovits:2004xu} \cite{Metsaev:1998it} \cite{Rahmfeld:1998zn} \cite{Park:1998un}, vertex operators and superstring amplitude computations in $\rm AdS$ are widely unexplored \cite{Berkovits:2022ivl}. 

In the string theory side, the difficulty can be tracked to the presence of Ramond-Ramond (R-R) flux in the worldsheet action. The R-R fields break the holomorphic/\allowbreak anti-holomorphic factorization of the worldsheet theory. As a result, the powerful complex methods of two-dimensional CFT \cite{DiFrancesco:1997nk} cannot be applied in a straightforward manner, and progress turned out to be rather slow in this direction since the advent of the $\rm AdS/CFT$ correspondence \cite{Maldacena:1997re}.

On the other hand, there exists an $\rm AdS$ target-space where holomorphic/anti-holomorphic factorization of the worldsheet theory is still preserved. This is the case for the Type IIB superstring in an $\rm AdS_3 \times S^3$ background in the absence of R-R fields, i.e., with pure Neveu-Schwarz-Neveu-Schwarz (NS-NS) self-dual three-form flux turned on. The holomorphic structure helps in the tractability of the theory. As a consequence, this particular example fits well to be a primary candidate for the understanding of quantitative features of covariant descriptions of the superstring in $\rm AdS$. The latter remark will be further explored in this work.

A spacetime covariant formulation of the Type IIB superstring in $\rm AdS_3 \times S^3$ is given by the hybrid formalism \cite{Berkovits:1994vy} \cite{Berkovits:1999im}. The sigma-model action of the hybrid string has the supergroup $\rm PSU(1,1|2)$ as the target-superspace and it can accommodate a mixture of both NS-NS and R-R constant three-form flux. In the pure NS-NS case, the worldsheet theory is given by a ${\rm PSU(1,1|2)}_k$ WZW model where $k$ labels the amount of NS-NS flux and is quantized \cite{Berkovits:1999im}. Along with the worldsheet action, the hybrid formalism enjoys a small $\mathcal{N}=4$ superconformal symmetry, and scattering amplitudes are computed according to the $\mathcal{N}=4$ topological prescription \cite{Berkovits:1994vy}.

In this work, we will make use of the pure NS-NS hybrid formalism in $\rm AdS_3 \times S^3$ with $k$ units of three-form flux and compute a three-point amplitude for half-BPS vertex operators inserted at a position $\mathbf{x}$ on the $\rm AdS_3$ boundary in a manifestly $\rm PSU(1,1|2)$-covariant fashion, i.e., using the spacetime supersymmetric worldsheet variables of the hybrid description. This will be done after defining curved worldsheet fields by making use of the vielbein field
\begin{align*}
{E_A}^B(\mathbf{x}) & = \dt_A^{B} + \mathbf{x} {f_{+ \, A}}^B - 2 \mathbf{x}^2 \eta_{A+} \dt^B_+ \,,
\end{align*}
where $A \in {\rm PSU(1,1|2)}$ Lie-superalgebra. 

The vertex operators depend on a fermionic coordinate $\ta^{\al}$. As an outcome, we will show that integrating out the fermions $\ta^{\al}(\mathbf{x})$ in the path integral gives rise to ${E_A}^B(\mathbf{x})$, which encodes that the conformal group on the boundary corresponds to the symmetry group of the $\rm AdS_3$ bulk \cite{Witten:1998qj}. Specifically, the fermionic zero-mode integration takes the following form
\begin{align*}
&\int d^4 \ta \, \ta^{\al}(\mathbf{x}_4) \ta^{\bt} (\mathbf{x}_3) \ta^{\g} (\mathbf{x}_2) \ta^{\dt}(\mathbf{x}_1) \nn \\
& \qquad = \ep^{\rho \sg \mu \nu} {E_{\rho 1}}^{\al 1}(-\mathbf{x}_4) {E_{\sg 1}}^{\bt 1}(-\mathbf{x}_3) {E_{\mu 1}}^{ \g 1}(-\mathbf{x}_2) {E_{\nu 1}}^{ \dt 1}(-\mathbf{x}_1) \,.
\end{align*}

Since spacetime supersymmetric superstring scattering amplitudes in curved backgrounds have been hardly ever investigated, this construction can have some important applications. In the first place, it provides intuition for what happens after the worldsheet fermions are integrated out in a general $\rm AdS$ background amplitude computation. Secondly, it gives insights about what the correct amplitude prescription in the more interesting case of $\rm AdS_5 \times S^5$ target-space might be. There have been significant works over the last years on trying to understand superstring vertex operators \cite{Berkovits:2019rwq} \cite{Fleury:2021ieo}, and the correct amplitude measure for the fermionic fields $\ta^{\al}$ in the $\rm AdS_5 \times S^5$ pure spinor formalism \cite{Berkovits:2008ga} \cite{Berkovits:2019ulm}.

Vertex operators for the pure R-R and pure NS-NS $\rm AdS_3 $ backgrounds were studied in refs.~\cite{Dolan:1999dc} and \cite{Langham:2000ac} in the hybrid formalism. In addition, for the pure R-R flux case, a three-graviton amplitude in the $\rm AdS_3$ bulk was calculated in ref.~\cite{Bobkov:2002bx}. For $k=1$ units of NS-NS flux, or tensionless point, superstring correlation functions for vertex operators inserted on the $\rm AdS_3$ boundary were computed in \cite{Dei:2020zui} \cite{Dei:2023ivl} and shown to match with the dual two-dimensional CFT \cite{Eberhardt:2018ouy}. Moreover, using a near boundary limit --- which is exact at $k=1$ --- these correlators were also analyzed in refs.~\cite{Knighton:2023mhq} \cite{Knighton:2024qxd} for the bosonic string and in ref.~\cite{Sriprachyakul:2024gyl} for the Ramond-Neveu-Schwarz (RNS) superstring.

There have been major advancements regarding string theory three-point functions in $\rm AdS_3$ with $k$ units of pure NS-NS flux and with vertex operators inserted in $\rm \pd AdS_3$. These advancements stem from both the RNS superstring formalism as well as from the bosonic $\rm SL(2 ,\mathbb{R})$ WZW model \cite{Maldacena:2001km} \cite{Gaberdiel:2007vu} \cite{Cardona:2009hk} \cite{Teschner:1999ug} \cite{Sriprachyakul:2024gyl}. More recently, a closed formula, in terms of an integral expression, has been derived for three-point functions in the $\rm SL(2, \mathbb{R})$ WZW model involving arbitrary spectrally flowed states \cite{Dei:2021xgh} \cite{Bufalini:2022toj}.

The hybrid formalism for the superstring offers a significant advantage over the RNS formalism by enabling quantization while manifestly preserving some of the spacetime supersymmetry (SUSY) \cite{Berkovits:1996bf}, and it eliminates the need for picture-changing or spin fields. Unlike the RNS formalism, the scattering amplitude prescription in the hybrid description is topological \cite{Berkovits:1994vy}, so that there is no need for summing over spin structures as well. Moreover, as opposed to the Green-Schwarz action, the hybrid action is quadratic in a flat background. 

The content of this paper is the following. In Section \ref{sechybridinAdS3}, we review the hybrid formalism in $\rm AdS_3 \times S^3$ background with pure NS-NS three-form flux together with spelling out our conventions for the worldsheet theory. In Section \ref{AdS3vertexsec}, we solve the physical state conditions and define half-BPS vertex operators in terms of a fermionic zero-mode coordinate $\ta^{\al}$. In Section \ref{seccurvedAdS3}, we perform a similarity transformation along the boundary direction and define the worldsheet fields and vertex operators depending on $\mathbf{x} \in {\rm \pd AdS_3}$ while introducing the vielbein field ${E_A}^B(\mathbf{x})$. In Section \ref{secthreepointhyb}, we compute a $\rm PSU(1,1|2)$-covariant three-point amplitude for vertex operators inserted on the $\rm AdS_3$ boundary and show that integrating out the fermionic fields $\ta^{\al}$ imply the appearance of ${E_A}^B(\mathbf{x})$ in the kinematic factor. In Section \ref{secRNSdescription}, we further confirm our results by comparing with the RNS formalism, and draw our conclusion in Section \ref{secconc}. At last, there exist a couple of appendices responsible for giving additional details.

\section{\boldmath Hybrid formalism in an $\rm AdS_3 \times S^3 $ background} \label{sechybridinAdS3}

In this section, we review the hybrid formalism in $\rm AdS_3 \times S^3$ with pure NS-NS self-dual three-form flux while defining our notation for the worldsheet theory. 

\subsection{Worldsheet action} 

The hybrid description \cite{Berkovits:1994vy} \cite{Berkovits:1999im} of the superstring in $\rm AdS_3 \times S^3 \times \mathcal{M}_4$, where $\mathcal{M}_4$ is either $\rm K3$ or $\rm T^4$, can be divided into a ``compactification-independent'' and a ``compactification-dependent'' part. The compactification-independent sector describes $\rm AdS_3 \times S^3$. It consists in a ${\rm PSU(1,1|2)}_k$ WZW model together with a $c=28$ chiral boson $\rho$ and the $c=-26$ chiral boson $\sg$. The compactification-dependent sector is composed of a twisted $c=6$ $\mathcal{N}=2$ superconformal field theory (SCFT) describing the four-dimensional manifold $\mathcal{M}_4$. One also has the right-moving counterpart of each of these sectors. Since the worldsheet theory enjoys a holomorphic/anti-holomorphic factorization, the right-movers will mostly be ignored for simplicity and clarity of the presentation.

The worldsheet action for the hybrid superstring in $\rm AdS_3 \times S^3$ is given by
\begin{align}\label{NSNSaction}
S & =  \frac{1}{2} k \int d^2 z \, \sTr\big( g^{-1} \pd g g^{-1} \pb g \big) - \frac{i}{2} k \int_{\mathcal{B}} \sTr \big( g^{-1} d g g^{-1} d g g^{-1} d g \big) \nn \\
& + S_{\rho, \sg} + S_C \,,
\end{align}
where $S_C$ is the action for the compactification directions containing four bosons and four fermions. The latter is defined by the twisted $c=6$ $\mathcal{N}=2$ SCFT it describes.  $S_{\rho , \sg}$ is the action for the chiral bosons $\{\rho ,\sg \}$, which is defined by the following OPE's for these fields
\begin{subequations}\label{chiralbosons}
\begin{align}
\rho(y) \rho (z) & \sim - \log(y-z) \,, \\
\sg(y) \sg(z) & \sim - \log(y-z)\,.
\end{align}
\end{subequations}

The first line of eq.~\eqref{NSNSaction} describes a ${\rm PSU(1,1|2)}_k$ WZW model. As a $\rm PSU(1,1|2)$ representative, one can take the group element
\begin{align}\label{groupAdS}
g &  = e^{Z^A T_A}\,,& Z^A&=\{ \ta^{\al j}, x^{\ab}\}\,,
\end{align}
where $A=\{\al j, \ab\}$ is a tangent space index and labels the supercoordinates, and $T_A$ are the generators of $\rm PSU(1,1|2)$ Lie superalgebra. The algebra generators satisfy the commutation relations
\begin{align}
[T_A,T_B\} &= {f_{A B}}^C T_C\,,& [T_A, T_B \} & = T_A T_B - (-)^{|A||B|} T_B T_A \,, 
\end{align}
whose structure constants are given by
\begin{align}\label{PSUstructureconst}
{f_{\al j \, \bt k}}^{\ab} & = i \sqrt{2} \ep_{jk} \sg^{\ab}_{\al \bt} \,,& {f_{\ab \, \al j}}^{\bt k} & = i \sqrt{2} \sg_{\ab \al \g} \dth^{\g \bt} \dt_j^k \,,& {f_{\ab \, \bb}}^{\cb} & = \sqrt{2} ({\sg_{\ab \bb}}^{\cb})_{\al \bt} \dth^{\al \bt} \,,
\end{align}
and where
\begin{align}
(\sg_{\ab \bb \cb})_{\al \bt} & = \frac{i}{3!} (\sg_{[\ab} \sg_{\bb} \sg_{\cb]})_{\al \bt} \,,& \dth^{\al \bt} &= 2 \sqrt{2} (\sg^{012})^{\al \bt}\,.
\end{align}
Note that in the equation above we anti-symmetrize with square brackets and without dividing by the number of terms. In our notation, $\ab=\{0$ to $5\}$ is an $\rm SO(1,5)$ vector index, $\al=\{1$ to $4\}$ is a fundamental $\rm SU(4)$ index and $j=\{1,2\}$ is an $\rm SU(2)$ index.

The four by four anti-symmetric matrices $\sg_{\ab \al \bt}$ are the $\rm SO(1,5)$ Pauli matrices which obey the Dirac algebra
\begin{align}
\sg^{\ab \al \bt} \sg^{\bb}_{\al \g} + \sg^{\bb \al \bt}\sg^{\ab}_{\al \g} &= \eta^{\ab \bb} \dt^\bt_\g \,,& \sg^{\ab \al \bt} & = \frac{1}{2} \ep^{\al \bt \g \dt} \sg^{\ab}_{\g \dt}\,,
\end{align}
where $\eta_{\ab \bb}$ is the usual mostly plus metric of the six-dimensional flat Minkowski background. In addition, the symbol $\dth_{\al \bt}$ and its inverse $\dth^{\al \bt}$ satisfy some interesting properties, namely,
\begin{align}
 \sg^{a}_{\al \bt} & = (\dth \sg^{a} \dth)_{\al \bt} \,, & \sg^{a^\prime}_{\al \bt} & = - (\dth \sg^{a^\prime} \dth)_{\al \bt} \,,& \dth^{\al \bt} \dth_{\bt \g}& = \dt_{\g}^{\al}\,,
\end{align}
where we write $a = \{0,1,2\}$  for the $\rm AdS_3$ directions and we write $a^\prime = \{3,4,5\}$ for the $\rm S^3$ directions. Supplementary identities for the six-dimensional Pauli matrices are given in Appendix \ref{sigmas}. 

The action \eqref{NSNSaction} is invariant under global left and right $\rm PSU(1,1|2)$ transformations of $g$, i.e.,
\begin{align}
{\rm PSU(1,1|2)_L \times PSU(1,1|2)_R}\,.
\end{align}
In particular, for the pure NS-NS case we are considering, this symmetry is actually enhanced to a local $g(z) \times g(\zb)$ symmetry acting on $g$ as
\begin{align}\label{localsymmetry}
g \rightarrow g_L(z) g g^{-1}_R(\zb) \,,
\end{align}
where $g_L(z)$($g_R(\zb)$) can be any holomorphic (anti-holomorphic) map from the worldsheet to ${\rm PSU(1,1|2)}$.

The supertrace over the $\rm PSU(1,1|2)$ generators defines the metric
\begin{align}
\sTr(T_A T_B) & = \eta_{AB}\,,& \eta^{AB} \eta_{BC} & = \dt^A_C\,,
\end{align}
whose non-zero components are
\begin{align}
\sTr(T_{\ab} T_{\bb}) & = \eta_{\ab \bb} \,,& \sTr(T_{\al j} T_{\bt k}) & =\ep_{jk} \dth_{\al \bt}\,,
\end{align}
where $\ep_{12}=\ep^{21}=1$ is the anti-symmetric tensor and $\dth_{\al \bt} = 2 \sqrt{2} (\sg^{012})_{\al \bt} $ is the symmetric matrix which enables one to contract spinor indices in an $\rm SO(1,2) \times SO(3)$ invariant manner.

For an object $X_A$ transforming in the representation $A$, we raise and lower tangent-space indices according to $X^A = \eta^{AB}X_B$ and $X_A = \eta_{AB} X^B$. Of course, the same rules apply for the structure constants $f_{ABC}$, which are graded anti-symmetric in the 1-2 and 1-3 indices.

From the fundamental field $g$ appearing in the worldsheet action, we define the left-currents by
\begin{align}
dg g^{-1} & = J_L^A T_{A} \,,
\end{align}
and the right-currents by
\begin{align}
g^{-1} d g & = J^A_R T_{A} \,,
\end{align}
where we write $J_L^A = \{S_L^{\al j} , K_L^{\ab} \}$ and $J_R^A = \{ S_R^{\al j}, K_R^{\ab} \}$. Note that the left-currents are right-invariant and the right-currents are left-invariant under global $\rm PSU(1,1|2)$ transformations. Although somewhat confusing, the latter statement is in fact correct.

The enhanced symmetry \eqref{localsymmetry} of the WZW model imply that the $(1,0)$ left-currents are purely holomorphic and the $(0,1)$ right-currents anti-holomorphic, i. e.,
\begin{align}
\pb (\pd g g^{-1}) & = 0 \,,& \pd (g^{-1} \pb g)& =0 \,.
\end{align}
Therefore, for simplicity of the notation, we will just write $J^A_{L \, z}  = J^A$ and $J^A_{R \, \zb} = \overline{J}^A$, so that the components read\footnote{Writing the fermionic currents as $S_{\al j}$ and the bosonic ones as $K_{\ab}$ turns out to give a more transparent notation for the $\mathcal{N}=4$ generators that we define in Section \ref{superconformalhyb}.}
\begin{align}\label{leftandrightJs}
J_A & = \{S_{\al j}, K_{\ab}\} \,,& \overline{J}_A & = \{\overline{S}_{\al j} , \overline{K}_{\ab} \}\,.
\end{align}

In addition, from the worldsheet action \eqref{NSNSaction} and after rescaling the currents by $k^{-1}$ and $k \rightarrow 2k$, one can show that the current algebra between the left-currents is
\begin{align}\label{currentalg2}
J_A(y) J_B(z) & \sim - \frac{2k}{(y-z)^2} \eta_{AB} + \frac{1}{(y-z)} {f_{AB}}^C J_C \,.
\end{align}
The current algebra between the anti-holomorphic right-currents can be derived from \eqref{currentalg2} by using the symmetry of the worldsheet action \eqref{NSNSaction} under $z \leftrightarrow \zb$ and $g \leftrightarrow g^{-1}$.

\subsection{Superconformal generators} \label{superconformalhyb}

The hybrid superstring description in $\rm AdS_3 \times S^3$ enjoys a twisted $c=6$ $\mathcal{N}=2$ superconformal symmetry generated by \cite{Berkovits:1999im}
\begin{subequations}\label{AdSN=2}
\begin{align}
T & = T_{\rm PSU} - \frac{1}{2} \pd \rho \pd \rho - \frac{1}{2} \pd \sg \pd \sg + \frac{3}{2} \pd^2(\rho + i \sg) + T_C  \,, \label{stressT} \\
G^+ & = - \frac{1}{4k}  (S_1)^4 e^{-2 \rho -i \sg} -  \frac{1}{2k}\bigg( \frac{i}{2 \sqrt{2}} S_{\al 1} S_{\bt 1}K^{\al \bt}  +\dth^{\al \bt} S_{\al 1} \pd S_{\bt 1} \bigg) e^{-\rho} \nn \\
& + T_{\rm {PSU}} e^{i \sg} +  \big( \pd e^{-\rho -i \sg} , e^{\rho + 2 i \sg} \big) + G^+_C \,, \label{GplusAdS3} \\
G^- & = e^{-i\sg} + G^-_C \,, \\
J & = \pd (\rho + i \sg) + J_C \,,
\end{align}
\end{subequations}
where $(S_1)^4 = \frac{1}{24} \ep^{\al \bt \g \dt} S_{\al 1} S_{\bt 1} S_{\g 1} S_{\dt 1}$, $K^{\al \bt} = \sg^{\ab \al \bt} K_{\ab}$. 

The ${\rm PSU(1,1|2)}_k$ stress-tensor is given by
\begin{align}\label{PSUstressT}
T_{\rm PSU}& = - \frac{1}{4k} J_A J_B \eta^{AB} \nn \\
& = - \frac{1}{4k}  \big( K_{\ab} K_{\bb} \eta^{\ab \bb} + S_{\al j} S_{\bt k} \eta^{\al j \bt k} \big) \,,
\end{align}
and the generators $\{G^{\pm}_C, T_C\}$ obey a twisted $c=6$ $\mathcal{N}=2$ superconformal algebra (SCA) for the compactification directions and have no poles with the $\{\rho, \sg\}$-ghosts and no poles with the matter currents.

Eqs.~\eqref{AdSN=2} need to be supplemented with a normal-ordering prescription for the ${\rm PSU(1,1|2)}_k$ currents. The normal-ordering is not commutative. We normal-order the currents according to
\begin{subequations}
\begin{align}
(J_A J_B)(z) & = \oint dy \, (y-z)^{-1}J_A (y) J_B(z) \,, \label{normalord}\\
J_AJ_B &= (-)^{|A||B|}J_BJ_A + {f_{AB}}^C \pd J_C\,.
\end{align}
\end{subequations}
The normal-ordering is also not associative. In our conventions, we normal-order from right to left so that $J_AJ_BJ_C = (J_A(J_BJ_C))$. 

Consequently, the ordering is not important in $T_{\rm PSU}$ because of the contraction with the metric, but it is important in the second term of the supercurrent $G^+$. In terms of modes, the normal-ordering is in agreement with the property
\begin{align}
(J_AJ_B)_0 V = (-)^{|A||B|} \nabla_B \nabla_A V\,,
\end{align}
for a $\rm PSU(1,1|2)$ primary field $V$ and with $\nabla_A$ the zero-mode of $J_A$.

The $c=6$ $\mathcal{N}=2$ SCA \eqref{AdSN=2} can be readily verified from the OPEs \eqref{chiralbosons} and \eqref{currentalg2}. The more complicated properties to check come from the supercurrent $G^+$. As shown in ref.~\cite{Berkovits:1999im}, one way to fix the form of the superconformal generator $G^+$ is by demanding the naive generalization from flat to curved space to be invariant under the ``non-standard'' spacetime supersymmetries generated by $Q_{\al 2}$. Another involved consistency condition to prove of the algebra generators \eqref{AdSN=2} is checking that the OPE of $G^+$ with itself is regular.

Instead of demonstrating term by term that $G^+$ commutes with $Q_{\al 2}$ and that $G^+(y)G^+(z) \sim 0$, we take a simpler route. Note that it is possible to write the supercurrent as
\begin{align}\label{Gplusid}
G^+ & = - \frac{1}{4k}  \frac{1}{24} \ep^{\al \bt \g \dt} Q_{\al 2} Q_{\bt 2} Q_{\g 2} Q_{\dt 2} e^{2 \rho + 3 i \sg} + G^+_C\,,
\end{align}
which makes manifest its invariance under the non-standard SUSYs generated by the charge
\begin{align}\label{nonstandardSUSY}
Q_{\al 2} & = \oint \big( S_{\al 1} e^{-\rho -i \sg} + S_{\al 2} \big) \,,
\end{align}
and also makes manifest the nilpotence property. Identity \eqref{Gplusid} is proved in Appendix \ref{Gplusidapp}.

In the hybrid formalism, the standard spacetime supersymmetry generator is
\begin{align}\label{standardSUSY}
Q_{\al 1} & = \oint S_{\al 1} \,,
\end{align}
and so we have the desired spacetime SUSY algebra
\begin{align}\label{spacetimeSUSYAdS}
\{Q_{\al j}, Q_{\bt k} \} & = {f_{\al j \, \bt k}}^{\ab } \oint K_{\ab} \,.
\end{align}
When mapped to the RNS description, $Q_{\al 1}$ and $Q_{\al 2}$ correspond to the spacetime SUSY generators in the $-\frac{1}{2}$ and $\frac{1}{2}$ picture, respectively. 

Let us emphasize that the reason for calling $Q_{\al 1}$ as the standard SUSY comes from the fact that it is ghost-independent, and so acts in a similar form as the supersymmetry generator of conventional superspace descriptions \cite{Howe:1983fr}. 

In what follows, we will write the zero-modes of the ${\rm PSU(1,1|2)}_k$ currents as $\nabla_A = \oint J_A$. More specifically, we define
\begin{align}\label{nablanotation}
\nabla_{\ab} & = \oint K_{\ab} \,,& \nabla_{\al j} & = \oint S_{\al j}\,.
\end{align}
This notation is convenient, since half the spacetime supersymmetries in the hybrid superstring $Q_{\al j}$ act different than the zero modes of the ${\rm PSU(1,1|2)}_k$ SUSY currents $S_{\al j}$. The latter is a consequence of the presence of the $\{ \rho, \sg\}$-ghosts in the four SUSYs $Q_{\al 2}$ of eq.~\eqref{nonstandardSUSY}.

Any twisted $c=6$ $\mathcal{N}=2$ SCA can be extended to a twisted small $c=6$ $\mathcal{N}=4$ SCA \cite{Berkovits:1994vy}. In addition to the generators \eqref{AdSN=2}, the remaining $\mathcal{N}=4$ generators of the hybrid formalism take the form
\begin{subequations}
\begin{align}
\widetilde{G}^+ & = e^{\rho} J^{++}_C - e^{\rho + i \sg} \widetilde{G}^+_C \,, \label{tildeGplus} \\ 
\widetilde{G}^- & = \bigg[ - \frac{1}{4k} (S_1)^4 e^{-3 \rho - 2 i \sg} + \frac{1}{2k} \bigg( \frac{i}{2 \sqrt{2}} S_{\al 1} S_{\bt 1} K^{\al \bt} + \dth^{\al \bt} S_{\al 1} \pd S_{\bt 1} \bigg) e^{-2 \rho -i \sg} \nn \\
& + T_{\rm PSU} e^{-\rho} - \big( \pd e^{-\rho -i \sg}, e^{i \sg} \big) \bigg] J^{--}_C + e^{- \rho -i \sg} \widetilde{G}^-_C \,, \\
J^{++} & =  - e^{\rho + i \sg} J^{++}_C \,, \\
J^{--} & = e^{-\rho - i \sg} J^{--}_C\,.
\end{align}
\end{subequations}
The generators $\{G^{\pm}_C, T_C\}$ together with $\{\widetilde{G}^{\pm}_C, J^{\pm \pm}_C\}$ form a twisted small $c=6$ $\mathcal{N}=4$ SCA for the compactification directions which has no poles with the $\{\rho ,\sg\}$-ghosts and no poles with the matter currents. Their explicit form is not needed in this work.

Remember that we are only discussing the holomorphic part, and hence one also has a right-moving twisted small $c=6$ $\mathcal{N}=4$ SCA. We display our conventions for the twisted $\mathcal{N}=2$ SCA and twisted small $\mathcal{N}=4$ SCA in Appendix \ref{N=4algApp}.

\subsection{Physical state conditions} \label{hybridphysical}

Physical states $\mathcal{V}$ of the hybrid superstring are defined to satisfy the following constraints \cite{Berkovits:1999im}
\begin{align}\label{hybridconstraints}
G^+_0 \widetilde{G}^+_0 \mathcal{V} & = 0 \,,& G^-_0 \mathcal{V} & = \widetilde{G}^-_0 \mathcal{V} = T_0 \mathcal{V} = J_0 \mathcal{V} = 0  \,,
\end{align}
and the state $\mathcal{V}$ is determined up to the gauge transformation
\begin{align}\label{gaugetransf}
\dt \mathcal{V} & = G^+_0 \Lambda + \widetilde{G}^+_0 \Omega + \widetilde{G}^-_0 \widetilde{G}^+_0 \Sigma \,,
\end{align}
where $\{\Lambda, \Omega \}$ are annihilated by $\{G^-_0,\widetilde{G}^-_0, T_0\}$ and $\Sigma$ is annihilated by $\{ G^-_0, T_0\}$. For a holomorphic operator $\mathcal{O}$ of conformal weight $h$, the notation $\mathcal{O}_n$ means the pole of order $n+h$.

Let us pause and comment about our gauge-fixing conditions. The first equation in \eqref{hybridconstraints} can be translated to the standard physical state condition of the RNS formalism $Q_{\rm RNS} V_{\rm RNS} = 0$, where $V_{\rm RNS}$ lives in the small hilbert space and is related to $\mathcal{V}$ as $\mathcal{V}= \xi V_{\rm RNS}$ (see Section \ref{secRNSdescription}). The constraint $T_0 \mathcal{V}=0$ is the usual mass-shell condition in string theory. When translated to RNS, the constraint $J_0 \mathcal{V}=0$ is equivalent as saying that the ghost- minus the picture-number of $V_{\rm RNS}$ is equal to one, as always happens for a physical RNS state \cite{Friedan:1985ge}. 

The additional constraints $G^-_0 \mathcal{V} = \widetilde{G}^-_0 \mathcal{V}=0$ in eqs.~\eqref{hybridconstraints} are convenient to further eliminate auxiliary degrees of freedom and imply a covariant gauge choice, e.g., they are equivalent to the Lorenz gauge condition for the open string sector \cite{Berkovits:1999im} \cite{Wess:1992cp} \cite{Berkovits:1997zd} \cite{Benakli:2021jxs}. As we will see in Section \ref{vertexAdS3}, it is also possible to define a Lorenz-type gauge in the $\rm AdS_3 \times S^3$ hybrid formalism which turns out to be suited for performing amplitude computations.

Note that in this formalism one of the candidates for the integrated vertex operator takes the form
\begin{align}\label{intvertexAdS3}
\int G^+_0 G^-_{-1}  \mathcal{V} \,,
\end{align}
which corresponds to a vertex operator in the same picture as $\mathcal{V}$ when translating to the RNS language \cite{Berkovits:1999im}. However, the important difference, when compared to the RNS formalism, is that $\mathcal{V}$ carries states both from the Ramond and Neveu-Schwarz sectors. In the RNS description, the Ramond states carry half-integer picture and the NS states carry integer picture. In fact, this is a crucial feature of the hybrid formalism. It treats Ramond and Neveu-Schwarz sectors in the same footing, since it only uses worldsheet variables of integer conformal weight.

\subsection{Amplitude prescription} \label{treeampprescription}

The prescription to compute $n$-point tree-level scattering amplitudes is given by
\begin{align}\label{npointhybrid}
\mathcal{A}_n & = \bigg\langle \mathcal{V}^{3}(z_3) \widetilde{G}^+_0\mathcal{V}^{(2)}(z_{2}) \bigg(\prod_{m=4}^{n} \int  dz_m \, G^-_{-1}G^+_0\mathcal{V}^{(m)}(z_m)\bigg) G^+_0 \mathcal{V}^{(1)}(z_1) \bigg\rangle \,, 
\end{align}
where $\mathcal{V}^{(n)}$ is the vertex operator satisfying the physical state conditions \eqref{hybridconstraints} and gauge transformations \eqref{gaugetransf}, and we are choosing $z_1=0$, $z_2 =1$ $z_3=\infty$ by $\rm SL(2, \mathbb{C})$ invariance. Note that the contribution from the right-movers is also being suppressed in $\mathcal{A}_n$.

In eq.~\eqref{npointhybrid}, the zero-mode integration over the fermions is done by generalizing the flat space prescription, i.e., 
\begin{align}\label{thetaintegration}
\int d^4 \ta& = \frac{1}{24} \ep^{\al \bt \g \dt} \nabla_{\dt 1} \nabla_{\g 1} \nabla_{\bt 1} \nabla_{\al 1} \nn \\
 & = (\nabla_1)^4\,,
\end{align}
where $\nabla_{\al 1}$ is the standard spacetime SUSY charge in $\rm AdS_3 \times S^3$, see eq.~\eqref{standardSUSY}. After integrating out the non-zero modes, the amplitude \eqref{npointhybrid} can always be expressed in terms of $\nabla_{\ab}$, the standard SUSY charge $\nabla_{\al 1}$ and the fermonic coordinate $\ta^{\al}$ (see eq.~\eqref{thetadefn} below). In particular, note that the measure \eqref{thetaintegration} is invariant under both $\nabla_{\ab}$ and $\nabla_{\al 1}$, so that the usual ``integration by parts'' is well defined.

The chiral bosons $\{\rho ,\sg\}$ carry a non-zero amount of background charge, as can be seen from eq.~\eqref{stressT}. Therefore, the tree-level amplitude is non-zero only when the path integral insertions contribute with the factor $e^{3\rho + 3 i\sg}$ in eq.~\eqref{npointhybrid}. In addition, since the compactification generators $\{T_C,G^{\pm}_C,J_C\}$ obey a twisted $c=6$ $\mathcal{N}=2$ SCA, one also needs the insertion of $J^{++}_C$. So that, in total, one gets $e^{3\rho + 3 i\sg}J ^{++}_C$ for the chiral bosons. In the amplitude \eqref{npointhybrid}, the factor of $J ^{++}_C$ comes from the term $\widetilde{G}^+_0 \mathcal{V}$. 

In this work, we will sometimes use definitions such as
\begin{subequations}
\begin{align}
(\ta^3)_{\al} & = \frac{1}{6} \ep_{\al \bt \g \dt} \ta^{\bt} \ta^{\g} \ta^{\dt} \,,& (\ta)^4 & = \frac{1}{24} \ep_{\al \bt \g \dt} \ta^{\al} \ta^{\bt} \ta^{\g} \ta^{\dt} \,, \\
(\nabla_1^3)^{\al} & = \frac{1}{6} \ep^{\al \bt \g \dt} \nabla_{\bt 1} \nabla_{\g 1} \nabla_{\dt 1} \,,& (\nabla_1)^4 & = \frac{1}{24} \ep^{\al \bt \g \dt} \nabla_{\al 1} \nabla_{\bt 1} \nabla_{\g 1} \nabla_{\dt 1} \,.
\end{align}
\end{subequations}

\section{\boldmath Vertex operators} \label{AdS3vertexsec}

This section deals with half-BPS vertex operators for the superstring in an $\rm AdS_3 \times S^3$ background with pure NS-NS three-form flux. After introducing the zero-mode variable $\ta^{\al}$, we will define the concept of a superfield in our superstring description for this background. Subsequently, the form of the half-BPS vertex operators will be determined by solving the constraints presented in Section \ref{hybridphysical}. 

For simplicity, we will consider vertex operators with no spectral flow \cite{Maldacena:2000hw}. Let us also emphasize that we will be working from the $\rm PSU(1,1|2)$ supergroup perspective in all stages of our development. For the readers not interested in the technical details, the gauge-fixed vertex operator is given in eq.~\eqref{vertexAdS34}.

\subsection{Superfields in $\rm AdS_3 \times S^3$} \label{AdS3superfield}

For the compactification-independent massless sector, the condition $T_0\mathcal{V}=0$ (see eqs.~\eqref{hybridconstraints}) imply that the vertex operator $\mathcal{V}$ in $\rm AdS_3 \times S^3$ transforms as a primary under the ${\rm PSU(1,1|2)}_k$ currents, i.e., 
\begin{align}\label{Vprimary}
J_A (y) \mathcal{V}(z) \sim (y-z)^{-1} \nabla_{A} \mathcal{V}(z) \,.
\end{align}
In particular, this implies that $\mathcal{V}$ has a pole with the fermionic current $S_{\al j}$. 

Since the standard SUSYs have the simple form \eqref{standardSUSY}, similar as in flat space, it is convenient to define $\mathcal{V}$ to be a superfield expanded in terms of a fermionic coordinate $\ta^{\al}$ which transforms as a Weyl spinor and is conjugate to $Q_{\al 1}$. More precisely, we define the superspace fermionic variable $\ta^{\al}$ by the property\footnote{Recall that in our notation $Q_{\al 1}=\nabla_{\al 1}$, see eqs.~\eqref{standardSUSY} and \eqref{nablanotation}.}
\begin{align}\label{thetadefn}
\nabla_{\al 1} \ta^{\bt} = \dt_{\al}^{\bt}\,.
\end{align}

Therefore, when we speak of a superfield in $\rm AdS_3 \times S^3$, we will be referring to a state which transforms as a primary under the ${\rm PSU(1,1|2)}_k$ currents and which has a finite expansion in terms of the fermionic coordinate $\ta^{\al}$. Note further that $\ta^{\al}$ is not trivially related to the group manifold coordinates $\ta^{\al j}$ in \eqref{groupAdS}, i.e., $\ta^{\al} \neq \ta^{\al 1}$ and $\ta^{\al } \neq \ta^{\al 2}$. 

From the definition \eqref{thetadefn}, we deduce that the remaining zero-modes of the ${\rm PSU(1,1|2)}_k$ currents satisfy
\begin{subequations}\label{thetadefn2}
\begin{align}
\nabla_{\ab} \ta^{\al} & = {f_{\bt 1 \, \ab}}^{\al 1} \ta^{\bt} \,, \label{thetadefn3} \\
\nabla_{\al 2} \ta^{\bt} & = \frac{1}{2} {f_{\al 2 \, \g 1}}^{\ab} {f_{\dt 1 \, \ab}}^{\bt 1} \ta^{\g } \ta^{\dt } \,.
\end{align}
\end{subequations}
Since $\mathcal{V}=\mathcal{V}(\ta)$, eq.~\eqref{thetadefn} also implies that the component fields of $\mathcal{V}$ are annihilated by $Q_{\al 1} = \nabla_{\al 1}$. In addition, we also have the expected property $T_0 \ta^{\al}  =0$, as can be easily checked.

In the flat background hybrid formalism, the vertex operator $\mathcal{V}$ for the massless compactification-independent sector is a superfield depending on a fermionic coordinate $\ta^{\al}$ and $\mathcal{V}$ has a simple pole with the standard spacetime supersymmetry current $p_{\al}$, since
\begin{align}\label{freefermions}
p_{\al}(y) \ta^{\bt}(z) & \sim \dt_{\al}^{\bt}(y-z)^{-1} \,,
\end{align}
where $\{p_{\al},\ta^{\bt}\}$ are holomorphic fermionic fundamental worldsheet fields of conformal weight one and zero, respectively. Therefore, one can view the definition \eqref{thetadefn} as a consequence of \eqref{Vprimary} and the generalization of the definition of $\mathcal{V}$ from the flat to the curved $\rm AdS_3 \times S^3$ spacetime. With the difference that $\ta^{\al}$ is not (a priori) a fundamental worldsheet coordinate in our description in terms of $g \in {\rm PSU(1,1|2)}$ in \eqref{NSNSaction}. Nevertheless, it is consistent to think of $\ta^{\al}$ as a fermionic zero-mode in the supergroup description and satisfying properties \eqref{thetadefn} and \eqref{thetadefn2}.

Regardless of that, it turns out that in a pure NS-NS $\rm AdS_3 \times S^3$ background the coordinate $\ta^{\al}$ can be viewed as a fundamental holomorphic worldsheet field. This hinges on the fact that the ${\rm PSU(1,1|2)}_k$ current algebra \eqref{currentalg2} can be expressed in terms of a ${\rm SU(1,1)}_{k+2} \times {\rm SU(2)}_{k-2}$ current algebra plus the eight free fermions \eqref{freefermions} \cite{Berkovits:1999im}.

Indeed, let $\ab = \{a, a^\prime\}$ where $a=\{0,1,2\}$ and $a^\prime=\{3,4,5\}$ label the $\rm AdS_3$ and $\rm S^3$ directions, respectively. We denote the ${\rm SU(1,1)}_{k+2} \times {\rm SU(2)}_{k-2}$ currents by $\mathcal{J}_{\ab}$.  The ${\rm SU(1,1)}_{k+2}$ current algebra reads
\begin{align}\label{RNScurrents1}
\mathcal{J}_a(y) \mathcal{J}_{b}(z) & \sim -\eta_{ab} \frac{2 (k+2)}{(y-z)^2} + \frac{1}{(y-z)} {f_{a \, b}}^c\mathcal{J}_c \,,
\end{align}
and the ${\rm SU(2)}_{k-2}$ reads
\begin{align}\label{RNScurrents2}
\mathcal{J}_{a^\prime}(y) \mathcal{J}_{b^\prime}(z) & \sim - \eta_{a^\prime b^\prime} \frac{2(k-2)}{(y-z)^2} + \frac{1}{(y-z)}{f_{a^\prime \, b^\prime}}^{c^\prime} \mathcal{J}_{c^\prime} \,,
\end{align}
where $f_{012}=f_{345}=-2$. Then, by defining
\begin{subequations}\label{RNScurrentalgebra}
\begin{align}
S_{\al 1} & = p_{\al} \,, \\
S_{\al 2} & = -2k \dth_{\al \bt} \pd \ta^{\bt} + {f_{\al 2 \, \bt 1}}^{\ab} \bigg( \mathcal{J}_{\ab} + \frac{1}{2} {f_{\ab \, \g 1}}^{\dt 1} p_{\dt} \ta^{\g}\bigg) \ta^{\bt}  \,, \\
K_{\ab} & = \mathcal{J}_{\ab} + {f_{\ab \, \al 1}}^{\bt 1}  p_{\bt} \ta^{\al} \,,
\end{align}
\end{subequations}
we recover the ${\rm PSU(1,1|2)}_k$ current algebra \eqref{currentalg2}, as we wanted to show. The bosonic curents $\mathcal{J}_{\ab}$ are the usual decoupled currents which appear in the RNS description \cite{Giveon:1998ns} \cite{Ferreira:2017pgt}. 

We should mention that in the hybrid description the eight free fermions $\{p_{\al},\ta^{\bt}\}$ come from a field redefinition involving the six $\psi^{\ab}$'s and the bosonized form of the $\{\bt, \g\}$-ghosts of the RNS formalism \cite{Berkovits:1999im}. As a corollary of this observation, one knows from the beginning that $\ta^{\al}(z)$ is holomorphic in a pure NS-NS $\rm AdS_3 \times S^3$ background.

Using eqs.~\eqref{RNScurrentalgebra}, one can readily check that the relations \eqref{thetadefn} and \eqref{thetadefn2} are reproduced. Except when comparing with RNS in Section \ref{secRNSdescription}, we will not use the explicit form of the currents \eqref{RNScurrentalgebra} in terms of the free fields $\{p_{\al},\ta^{\bt}\}$. This means that we will be working from the supergroup perspective and hence with the currents constructed from $g \in {\rm PSU(1,1|2)}$.  However, it will be assumed eq.~\eqref{thetadefn}, which naturally follows from the generalization of a ``superspace coordinate'' from the flat to the curved $\rm AdS_3 \times S^3$ background in the hybrid formalism. 

In this case, let us emphasize that $\ta^{\al}$ is a fermionic zero-mode (Schrödinger operator) which is responsible for building up our physical states in a covariant fashion. Therefore, it will not be necessary to know how it depends on $g(z, \zb)$ in this work, and so this reasoning should also generalize to $\rm AdS_3 \times S^3$ when turning on a constant R-R three-form flux \cite{CD25}.

\subsection{Vertex operators for the massless states}\label{vertexAdS3}

Now, we will determine the gauge-fixed half-BPS vertex operators by analyzing the physical state conditions of Section \ref{hybridphysical}. As usual, we concentrate on the holomorphic part of the theory. 

For the massless compactification-independent states (i.e., states of conformal weight zero at zero momentum) of the Type IIB superstring in $\rm AdS_3 \times S^3$, the condition that the vertex operator $\mathcal{V}$ should have no single poles with $J$ imply that it takes the form
\begin{align}\label{vertexAdS31}
\mathcal{V}& = \sum_n e^{n(\rho + i \sg)} V_n \,.
\end{align}

Demanding $\mathcal{V}$ to have no double poles or higher with $G^-$ and no double poles or higher with $\widetilde{G}^-$ imply that $V_n = 0$ for $n \geq 2$ and $V_n = 0$ for $n \leq -2$, respectively. Moreover, the condition $\widetilde{G}^-_0 \mathcal{V} =0$ also gives the following constraints for the remaining superfields $\{V_{-1}, V_0, V_1 \}$
\begin{subequations}\label{EOMsAdS3}
\begin{align}
\nabla_{\al 1} V_{-1} & = 0 \,, \label{EOMsAdS31}\\
\frac{i}{2\sqrt{2}} \sg^{\al \bt}_{\ab} \nabla_{\al 1} \nabla_{\bt 1} V_0 - \nabla_{\ab} V_{-1} & = 0 \,,\label{EOMsAdS32} \\
(\nabla_1)^4 V_1 & = 0 \,,\label{EOMsAdS33} \\
(\nabla_1^3)^{\al} V_1 + \Big( -i \sqrt{2} \nabla^{ \al \bt} \nabla_{\bt 1} + 2 \dth^{\al \bt} \nabla_{\bt 1} \Big)V_0 - 2 \dth^{\al \bt} \nabla_{\bt 2} V_{-1} & = 0\,,\label{EOMsAdS34} \\
\nabla^{\al \bt} \nabla_{\al 1} \nabla_{\bt 1} V_ 1 & = 0 \,. \label{EOMsAdS35}
\end{align}
\end{subequations}

Let us now determine what are the physical states by analyzing eqs.~\eqref{EOMsAdS3} together with the gauge transformations \eqref{gaugetransf} for the remaining superfields. From eq.~\eqref{EOMsAdS31} we learn that $V_{-1}$ has no components proportional to $\ta^{\al}$. By taking $\Sigma = 4k e^{\rho + i \sg} (\ta)^4 V_{-1}$ in \eqref{gaugetransf}, we see that $V_{-1}$ can be gauged away. Therefore, eq.~\eqref{EOMsAdS32} implies that $V_0 = v_0 + \ta^{\al} \chi_{\al 1}$ for some $\{v_0, \chi_{\al 1}\}$. Actually, the component $v_0$ can be removed by taking $\Omega = - e^{\rho} J^{++}_C v_0$ in the gauge transformations \eqref{gaugetransf}. Therefore, we conclude that one can gauge-fix $V_0$ to the form
\begin{align}
V_0 & = \ta^{\al} \chi_{\al 1} \,.
\end{align}

We now turn to analyze the components of the superfield $V_1$, the most important part of the vertex operator $\mathcal{V}$. Firstly, eq.~\eqref{EOMsAdS33} implies that $V_1$ has no $(\ta)^4$ component. Now, consider the gauge transformation given by
\begin{subequations}
\begin{align}
\Lambda & = 2\sqrt{2} k e^{2 \rho + i \sg} \xi \,, \\
\xi & = - \frac{i}{2} (\ta \sg_{\ab} \ta) \omega^{\ab} + i (\ta^3)_{\al} \tau^{\al} + (\ta)^4 \la \,,
\end{align}
\end{subequations}
for some $\{\omega^{\ab}, \tau^{\al}, \la \}$. From \eqref{gaugetransf}, one finds
\begin{align}\label{gaugetransfAdS3}
\dt V_1 & = - \frac{i}{2} \nabla^{\al \bt} \nabla_{\al 1} \nabla_{\bt 1} \xi \nn \\
& = \nabla^{\ab} \omega_{\ab} + \ta^{\al} \nabla_{\al \bt} \tau^{\bt} + \frac{i}{2} (\ta \sg^{\ab} \ta) \nabla_{\ab} \la\,.
\end{align}
Using the gauge parameter $\omega^{\ab}$, one can gauge away the first component of $V_1$. As a result, we can gauge-fix the superfield $V_1$ to the following form
\begin{align}\label{vertexAdS32}
V_1 & = \ta^{\al} \chi_{\al 2} + \frac{i}{2} (\ta \sg_{\ab} \ta) a^{\ab} - (\ta^3)_{\al} \psi^{\al 2} \,.
\end{align}

In addition, eq.~\eqref{EOMsAdS34} implies that $\psi^{\al 2}=(i \sqrt{2} \nabla^{\al \bt} -2 \dth^{\al \bt})\chi_{\bt 1}$. As a consequence, all the degrees of freedom are contained in the superfield $V_1$. In view of that and for later convenience, we define $V_1 = V$. Therefore, the gauge-fixed vertex operator \eqref{vertexAdS31} takes the form
\begin{align}\label{vertexAdS34}
\mathcal{V}& = e^{\rho + i \sg} V + V_0 \,,
\end{align}
where
\begin{subequations}\label{vertexAdS35}
\begin{align}
V & = \ta^{\al} \chi_{\al 2} + \frac{i}{2} (\ta \sg_{\ab} \ta) a^{\ab} - (\ta^3)_{\al} \psi^{\al 2} \,, \label{vertexAdS33}\\
V_0 & = \ta^{\al} \chi_{\al 1} \,.
\end{align}
\end{subequations}
Note further that the superfield $V$ satisfies the equation of motion $\nabla^{\al \bt} \nabla_{\al 1} \nabla_{\bt 1} V =0$ (see eq.~\eqref{EOMsAdS35}). 

All component fields obey the mass-shell condition $\nabla^{\ab}\nabla_{\ab}=0$. It is also convenient to define the gauge invariant ``fermions''
\begin{align}
\psi^{\al j} & = \ep^{jk} (i \sqrt{2} \nabla^{\al \bt} - 2 \dth^{\al \bt})\chi_{\bt k}\,. 
\end{align}
Note that these fermions satisfy the ``Dirac-like'' equation
\begin{align}
D_{\al \bt} \psi^{\bt j} & =0 \,,
\end{align}
in curved space, where $D_{\al \bt} = \sg^{\ab}_{\al \bt} D_{\ab}$. 

For an object $X_A$ transforming in the representation $A$ of $\rm PSU(1,1|2)$, we define
\begin{align}\label{covderdefn}
D_{\ab} X_B & = \nabla_{\ab} X_B - \frac{1}{2} {f_{\ab \, B}}^C X_C\,.
\end{align}
In fact, one can show that $D_{\al \bt} \psi^{\bt j} =\nabla_{\al \bt} \psi^{\bt j}$ by using the explicit form of the structure constants \eqref{PSUstructureconst}.

In terms of the RNS formalism language, the components of $V$ proportional to $(\ta \sg_{\ab} \ta)$ are states from the NS-sector and the components proportional to $\ta^{\al}$ and $(\ta^3)_{\al}$ are states from the R-sector.

Although we are only discussing the holomorphic part of the theory for simplicity, the identification of the equations of motion derived from the string constraints \eqref{hybridconstraints} with the supergravity field equations in $\rm AdS_3 \times S^3$ was elaborated in ref.~\cite{Langham:2000ac}.

Since the fermionic variables $\ta^{\al}$ are charged under the $\rm SL(2, \mathbb{R}) \times SU(2)$ bosonic subgroup of $\rm PSU(1,1|2)$, we can also relate the components of the superfield in \eqref{vertexAdS35} with the Maldacena-Ooguri vertex operators described in terms of the $\rm SL(2, \mathbb{R})$ and $\rm SU(2)$ quantum numbers \cite{Maldacena:2000hw} \cite{Maldacena:2001km}. We refer to Appendix \ref{MOvertices} for this description.

\section{Curved worldsheet fields} \label{seccurvedAdS3}

For the purpose of computing $\rm PSU(1,1|2)$-covariant superstring scattering amplitudes with vertex operators being functions of the spacetime boundary positions, we will introduce worldsheet fields depending on the boundary $\rm AdS_3$ coordinates $\mathbf{x}$. This will be done by performing a similarity transformation in the $\nabla_+ $ direction with parameter the complex coordinate $\mathbf{x}$ and where 
\begin{align}\label{translationvertex}
\nabla_{+} & = - \frac{i}{2} \oint (K_1 + i K_2)\,,
\end{align}
is the translation generator along the $\rm AdS_3$ boundary or, equivalently, in the dual CFT \cite{Giveon:1998ns}. Naturally, the vielbein field ${E_A}^B(\mathbf{x})$ will emerge in our description.

When writing a field $\mathcal{O}$ without any labels, it means that it only depends on the worldsheet coordinates $z$. Therefore, it is inserted in the position $\mathbf{x}=0$ on the boundary.\footnote{I would like to thank Lucas Martins and Dennis Zavaleta for discussions regarding this point.} For an operator function of any $\mathbf{x} \in \pd {\rm AdS}_3$, we will write $\mathcal{O}(\mathbf{x},z)$ --- or simply $\mathcal{O}(\mathbf{x})$ --- which is equivalent to $e^{\mathbf{x}\nabla_+} \mathcal{O} e^{-\mathbf{x} \nabla_+}$. As we will presently see, there are only a finite number of terms that contribute in this similarity transformation for our fundamental worldsheet variables.

Vertex operators translated by the generator \eqref{translationvertex} were used in refs.~\cite{Eberhardt:2019ywk} \cite{Dei:2020zui} to match worldsheet correlators at $k=1$ units of NS-NS flux with the dual two-dimensional CFT correlators \cite{Lunin:2000yv} \cite{Lunin:2001pw}.

\subsection{Similarity transformation and the vielbein}

Consider the holomorphic ${\rm PSU(1,1|2)}_k$ currents \eqref{leftandrightJs}, the effect of introducing dependence on the boundary $\rm AdS_3$ coordinates $\mathbf{x}$ is given by
\begin{align}
J_A(\mathbf{x},z)& = e^{\mathbf{x} \nabla_+} J_A(z) e^{-\mathbf{x} \nabla_+} \nn \\
& = J_A(z) + \mathbf{x} {f_{+ \, A}}^B J_B(z) + \frac{\mathbf{x}^2}{2} {f_{+ \, A}}^B {f_{+ \, B}}^C J_C(z)\,,
\end{align}
since
\begin{align}
{f_{+ \, A}}^B {f_{+ \, B}}^C =-4 \dt_+^C \eta_{A+}\,.
\end{align}
This means that we can write
\begin{align}
J_A (\mathbf{x},z) & = {E_A}^B(\mathbf{x}) J_B(z)\,,
\end{align}
where
\begin{align}\label{vielbeinfieldAdS3}
{E_A}^B(\mathbf{x}) & = \dt_A^{B} + \mathbf{x} {f_{+ \, A}}^B - 2 \mathbf{x}^2 \eta_{A+} \dt^B_+ \,,
\end{align}
and so the matrix ${E_A}^B(\mathbf{x})$ converts a flat worldsheet field to a curved one (in spacetime). 

Therefore, we will take our freedom and call the quadratic matrix ${E_A}^B(\mathbf{x})$ the vielbein field \cite{Wess:1992cp}. In particular, note from \eqref{vielbeinfieldAdS3} that ${E_A}^B(\mathbf{x})$ has a finite number of terms and at most quadratic in $\mathbf{x}$. Since $J_A = \{S_{\al j}, K_{\ab}\}$, in our usual notation, we simply write
\begin{align}
S_{\al j}( \mathbf{x}) & = {E_{\al j}}^{\bt k}(\mathbf{x}) S_{\bt k}\,,& K_{\ab} (\mathbf{x}) & = {E_{\ab}}^{\bb}(\mathbf{x}) K_{\bb} \,.
\end{align}

In supergravity descriptions, the vielbein field ${E_A}^B(\mathbf{x})$ carries a lower Einstein index and an upper Lorentz (or structure group) index \cite{Wess:1992cp}, and ${E_A}^B(\mathbf{x})$ is not written with an explicit spacetime dependence of $\mathbf{x}$. In this work, we will not differentiate between Einstein and Lorentz indices. However, this should yield no confusion, since for any object $\mathcal{O}$ depending on $\mathbf{x}$ we will explicit write $\mathcal{O}(\mathbf{x})$.  

The vertex operator \eqref{vertexAdS34} in the $\mathbf{x}$-basis $\mathcal{V}(\mathbf{x},z)$ is
\begin{align}\label{similaritytransfV}
\mathcal{V}(\mathbf{x},z) & = e^{\mathbf{x} \nabla_+}\mathcal{V}(z) e^{-\mathbf{x}\nabla_+} \,.
\end{align}
therefore, from \eqref{Vprimary}, the action of the spacetime dependent ${\rm PSU(1,1|2)}_k$ currents $J_A$ is given by
\begin{align}
J_{A}(\mathbf{x}_1, y) \mathcal{V}(\mathbf{x}_2, z) & \sim (y-z)^{-1} \bigg( \nabla_A\mathcal{V}  + \mathbf{x}_{12} {f_{+\, A}}^B \nabla_B\mathcal{V}  -2 \mathbf{x}_{12}^2 \eta_{A+} \nabla_+\mathcal{V} \bigg) (\mathbf{x}_2 ,z )\,,
\end{align}
where $\mathbf{x}_{12}= \mathbf{x}_1  - \mathbf{x}_2 $. In our formulas, we shall also write
\begin{align} 
J_{A}(\mathbf{x}_1, y) \mathcal{V}(\mathbf{x}_2, z) & \sim (y-z)^{-1} \big(\nabla_{A}(\mathbf{x}_{12})\mathcal{V} \big)(\mathbf{x}_2 ,z )\nn \\
& = (y-z)^{-1} {E_A}^B(\mathbf{x}_{12}) (\nabla_B \mathcal{V} )(\mathbf{x}_2 ,z )\,,
\end{align}
to simplify the notation.

\subsection{Curved fermionic coordinates}

As we discussed in the beginning of Section \ref{AdS3vertexsec}, the vertex $\mathcal{V}$ in eq.~\eqref{vertexAdS34} is a superfield in our superstring description, i.e., it is a function of the fermionic zero-mode variable $\ta^{\al}$. In order to compute amplitudes involving the vertex $\mathcal{V}$ inserted on the $\rm AdS_3$ boundary, we need to specify what are the analogues of the superspace coordinates $\ta^{\al}$ when we introduce dependence on the spacetime coordinate $\mathbf{x}$.

As before, in the $\mathbf{x}$-basis, we have that
\begin{align}
\ta^{\al} (\mathbf{x}) &= e^{\mathbf{x} \nabla_+} \ta^{\al} e^{- \mathbf{x} \nabla_+} \,,
\end{align}
and eq.~\eqref{thetadefn3} implies
\begin{align}
\ta^{\al} (\mathbf{x}) & = {E_{\bt 1}}^{\al 1} (- \mathbf{x}) \ta^{\bt} \nn \\
& = \big( \dt^{\al}_{\bt} + \mathbf{x} {f_{\bt 1 \, +}}^{\al 1} \big) \ta^{\bt}  \nn \\
& = \ta^{\al}  - \mathbf{x} i \sqrt{2} (\ta \sg_+\dth)^{\al}\,.
\end{align}
Therefore, from \eqref{thetadefn}, one finds that the action of the standard SUSYs on $\ta^{\al}(\mathbf{x})$ is
\begin{align}\label{thetadefn4}
\nabla_{\al 1} \ta^{\bt} (\mathbf{x}) & = {E_{\al 1}}^{\bt 1}(-\mathbf{x})\nn \\
& = \dt^{\bt}_{\al} - \mathbf{x} i \sqrt{2} (\sg_+ \dth)_{\al}^{\ \bt}\,,
\end{align}
where ${E_{\al 1}}^{\bt 1}(\mathbf{x})$ is the vielbein field of eq.~\eqref{vielbeinfieldAdS3}.

From the last property, together with eq.~\eqref{thetaintegration}, we then have determined a way to integrate the curved worldhsheet fermions $\ta^{\al} (\mathbf{x})$ in a tree-level amplitude computation. The answer is given in terms of the vielbein, namely,\footnote{ To perform calculations, it is actually easier to use the less condensed but more practical notation of a curved delta-function for the spinorial vielbein, i.e.,
\begin{align}
{E_{\al 1}}^{\bt 1}(-\mathbf{x}) & = \dt_{\al}^{\bt}(\mathbf{x})\,.
\end{align}}
\begin{align}\label{thetaintAdS}
&\int d^4 \ta \, \ta^{\al}(\mathbf{x}_4) \ta^{\bt} (\mathbf{x}_3) \ta^{\g} (\mathbf{x}_2) \ta^{\dt}(\mathbf{x}_1) \nn \\
& \qquad = \ep^{\rho \sg \mu \nu} {E_{\rho 1}}^{\al 1}(-\mathbf{x}_4) {E_{\sg 1}}^{\bt 1}(-\mathbf{x}_3) {E_{\mu 1}}^{ \g 1}(-\mathbf{x}_2) {E_{\nu 1}}^{ \dt 1}(-\mathbf{x}_1) \,.
\end{align}
If desired, the expression above can be explicitly evaluated using the definition \eqref{vielbeinfieldAdS3}, one finds
\begin{align}
&\int d^4 \ta \, \ta^{\al}(\mathbf{x}_4) \ta^{\bt} (\mathbf{x}_3) \ta^{\g} (\mathbf{x}_2) \ta^{\dt}(\mathbf{x}_1) \nn \\
& \qquad = \ep^{\al \bt \g \dt} + \frac{i}{\sqrt{2}} \Big( -\mathbf{x}_4 \dth^{\al [ \bt} \sg_+^{\g \dt]} + \mathbf{x}_3 \dth^{\bt[\al} \sg_+^{\g \dt]} - \mathbf{x}_2 \dth^{\g [ \al} \sg_+^{\bt \dt]} + \mathbf{x}_1 \dth^{\dt[\al} \sg_+^{\bt \g]} \Big)\nn \\
& \qquad - 2 \sg_+^{\al \bt} \sg_+^{\g \dt} \big( \mathbf{x}_1 \mathbf{x}_2 + \mathbf{x}_3 \mathbf{x}_4 \big) + 2 \sg_+^{\al \g} \sg_+^{\bt \dt} \big( \mathbf{x}_1 \mathbf{x}_3 + \mathbf{x}_2 \mathbf{x}_4 \big) -2 \sg_+^{\al \dt} \sg_+^{\bt \g} \big( \mathbf{x}_1 \mathbf{x}_4 + \mathbf{x}_2 \mathbf{x}_3 \big) \,,
\end{align}
hence, only terms up to quadratic-order in $\mathbf{x}$ appear when integrating out the fermionic zero-modes $\ta^{\al}(\mathbf{x})$'s.

\subsection{Some properties of the vielbein} \label{curvedsymb}

We have explicitly shown how flat worldsheet fields can be made dependent on the boundary $\rm AdS_3$ coordinates $\mathbf{x}$. One of the key ideas is the presence of the spacetime dependent matrix \eqref{vielbeinfieldAdS3}, which naturally appears in our superstring description after performing a similarity transformation in the direction $\nabla_+$ with parameter $\mathbf{x}$. 

For the purpose of carrying out computations, it is useful to state some of the identities satisfied by ${E_A}^B(\mathbf{x})$. One can show that
\begin{align}
{E_A}^B(\mathbf{x}) {E_B}^C(-\mathbf{x}) & =\dt_A^C\,.
\end{align}
and, note also
\begin{subequations}
\begin{align}
{E_{\al 1}}^{ \g 1}(-\mathbf{x}) \sg_{\ab \g \dt} {E_{\bt 1}}^{ \dt 1}(-\mathbf{x}) & = {E_{\ab}}^{ \bb} (\mathbf{x}) \sg_{\bb \al \bt}\,, \\
{E_{\g 1}}^{ \al 1}(-\mathbf{x}) \sg^{ \g \dt}_{\ab} {E_{\dt 1}}^{\bt 1}(-\mathbf{x}) & = {E_{\ab}}^{\bb} (-\mathbf{x}) \sg_{\bb}^{\al \bt}\,,
\end{align}
\end{subequations}
and that
\begin{align}
e^{\mathbf{x} \nabla_+} (\ta \sg_{\ab} \ta) e^{-\mathbf{x} \nabla_+} & = {E_{\ab}}^{\bb} (\mathbf{x}) (\ta \sg_{\bb} \ta) \,,
\end{align}
hence, ${E_A}^B(\mathbf{x})$ transforms a ``flat Pauli matrix'' to a ``curved one'' in spacetime, as is expected for a vielbein \cite{Wess:1992cp}.

In particular, the vielbein field with bosonic indices ${E_{\ab}}^{ \bb}(\mathbf{x})$ satisfy
\begin{subequations}
\begin{align}
{E_{\ab}}^{ \bb} (\mathbf{x}) & = {E^{\bb}}_{\ab} (-\mathbf{x})\,, \\
{E_{ \ab}}^{ \cb}(\mathbf{x}_i) {E_{ \bb}}^{ \db}(\mathbf{x}_j) \eta_{\cb \db} & = E_{\ab \bb}(\mathbf{x}_{ij}) \,, 
\end{align}
\end{subequations}
where we are denoting $E_{\ab \bb}(\mathbf{x}) = \eta_{\bb \cb} {E_{\ab}}^{\cb}(\mathbf{x})$.

We also have that
\begin{align}
[{E_{\ab}}^{ \cb}(\mathbf{x})\pd_{\cb}, {E_{\bb}}^{ \db}(\mathbf{x})\pd_{\db}]& = {c_{\ab \bb}}^{\cb} (\mathbf{x}) {E_{\cb}}^{ \db}(\mathbf{x}) \pd_{\db} \,,
\end{align}
where
\begin{align}
{c_{\ab \bb}}^{\cb} ( \mathbf{x}) & = \dt^{+}_{[\ab |}{f_{+ \, |\bb]}}^{\cb} + \mathbf{x} {f_{+ \, [\ab |}}^+ {f_{+ \, |\bb]}}^{\cb} - 2 \mathbf{x}^2 \eta_{+[\ab|} {f_{+ \, |\bb]}}^{\cb} \,,
\end{align}
and we used $\pd_{\ab} \mathbf{x} = \dt^+_{\ab}$.

As a consequence, one identifies
\begin{align}\label{mobiusgen}
{E_{ +}}^{+}\pd_+ & = \pd_+ \,,& {E_{3}}^{+} \pd_+& = - \mathbf{x} \pd_+ \,,& {E_{-}}^{+} \pd_+& = \mathbf{x}^2 \pd_+\,,
\end{align}
as the generators of infinitesimal two-dimensional conformal transformations. In effect, eqs.~\eqref{mobiusgen} highlight that the conformal group acting on the boundary corresponds to the symmetry group of the bulk $\rm AdS_3$ spacetime \cite{Witten:1998qj}. Hence, we found a standard property of the $\rm AdS/CFT$ correspondence folklore via a first-principles superstring theory calculation.\footnote{Recall that the conformal group in $\rm AdS_3$ is $\rm SO(2,2) \cong SU(1,1)_L \times SU(1,1)_R$ and we are only displaying the holomorphic part of the worldsheet theory.} This observation might give important hints towards the correct description of superstring vertex operators in $\rm AdS_5 \times S^5$ \cite{Berkovits:2008ga} \cite{Berkovits:2019ulm}.

For the purpose of computing scattering amplitudes, we also define the curved structure constants
\begin{align}\label{structofx2}
f_{\ab \bb \cb}(\mathbf{x}_1, \mathbf{x}_2, \mathbf{x}_3) & = {E_{\ab}}^{\db}(\mathbf{x}_1) {E_{\bb}}^{\eb}(\mathbf{x}_2) {E_{\cb}}^{\fb}(\mathbf{x}_3) f_{\db \eb \fb} \,,
\end{align}
which only depend on the distance $ \mathbf{x}_{ij}= \mathbf{x}_i - \mathbf{x}_j$, as can be easily seen from the explicit expression
\begin{align}\label{structofx}
f_{\ab \bb \cb} (\mathbf{x}_1 , \mathbf{x}_2, \mathbf{x}_3) &= f_{\ab \bb \cb} + 4 \Big( \mathbf{x}_{12} \eta_{\cb +} \eta_{\ab \bb} - \mathbf{x}_{13} \eta_{\bb +} \eta_{\ab \cb} + \mathbf{x}_{23} \eta_{\ab +} \eta_{\bb \cb} \Big) \nn \\
&  -2 \Big( \mathbf{x}_{12} \mathbf{x}_{13} \eta_{\ab +} f_{+ \bb \cb } - \mathbf{x}_{12} \mathbf{x}_{23} \eta_{\bb +} f_{+ \cb \ab} + \mathbf{x}_{13} \mathbf{x}_{23} \eta_{\cb +} f_{+ \ab \bb} \Big) \nn \\
&  + 8 \mathbf{x}_{12} \mathbf{x}_{13} \mathbf{x}_{23} \eta_{\ab +} \eta_{\bb +} \eta_{\cb +} \,.
\end{align}
Furthermore, this means that the curved structure constants \eqref{structofx2} are invariant under a constant shift of $\{\mathbf{x}_1,\mathbf{x}_2,\mathbf{x}_3\}$, i.e., they satisfy
\begin{align}
f_{\ab \bb \cb}(\mathbf{x}_1, \mathbf{x}_2, \mathbf{x}_3) & = f_{\ab \bb \cb}(\mathbf{x}_{14} , \mathbf{x}_{24} , \mathbf{x}_{34} ) \,,
\end{align}
for any $ \mathbf{x}_4$.

\section{Amplitude computation}\label{secthreepointhyb}

In Section \ref{sechybridinAdS3}, we introduced the worldsheet action for the $\rm AdS_3 \times S^3$ hybrid formalism together with the OPEs satisfied by the fundamental fields: the $\{\rho,\sg\}$-ghosts and the ${\rm PSU(1,1|2)}_k$ currents. We also defined constraints that determine the physical states in a suitable gauge choice and wrote a tree-level scattering amplitude prescription. In particular, the fermionic measure of integration was described in terms of the standard spacetime SUSYs $\nabla_{\al 1}$.

In Sections \ref{AdS3vertexsec} and \ref{seccurvedAdS3}, after introducing the zero-mode fermionic coordinate $\ta^{\al}$, we determined the gauge-fixed vertex operators for the half-BPS states. Additionally, we showed how vertex operators inserted at $\mathbf{x}=0$ can be translated to an arbitrary position $\mathbf{x}$ on the $\rm AdS_3$ boundary by the means of a similarity transformation and using the vielbein ${E_A}^B(\mathbf{x})$.

That being said, we have collected enough information to calculate tree-level $\rm PSU(1,1|2)$-covariant scattering amplitudes for half-BPS vertex operators in $\rm AdS_3$. For this reason, the content of this section is to exemplify how these tools can be used in practice by computing a three-point amplitude and highlighting some new features present in this covariant approach.

\subsection{Three-point amplitude in $\rm AdS_3$} 

Following the prescription \eqref{npointhybrid}, the three-point amplitude for the half-BPS vertex operator $\mathcal{V}$ in \eqref{vertexAdS34} and \eqref{similaritytransfV} is given by
\begin{align}\label{3pointAdS}
\mathcal{A}_3 & = \bigg\langle  \mathcal{V}^{(3)}(\mathbf{x}_3,z_3) \big( \widetilde{G}^+_0 \mathcal{V}^{(2)}\big)(\mathbf{x}_2, z_2) \big( G^+_0 \mathcal{V}^{(1)}\big)(\mathbf{x}_1, z_1) \bigg\rangle\,,
\end{align}
where
\begin{subequations}\label{3pointterms}
\begin{align}
\mathcal{V}(\mathbf{x},z) &  =e^{\rho+ i \sg} V (\mathbf{x}, z)  \label{3pointterms1}\,,\\
\big( \widetilde{G}^+_0 \mathcal{V}\big)(\mathbf{x}, z) & =e^{2 \rho + i \sg } J^{++}_C V (\mathbf{x},z) \label{3pointterms2} \,, \\
\big( G^+_0 \mathcal{V}\big)(\mathbf{x}, z)&  = - \frac{1}{2k} e^{i\sg} \bigg[ \frac{i}{2\sqrt{2}} \bigg( K^{\al \bt} \nabla_{\al 1} \nabla_{\bt 1} V  + 2  S_{\al 1}\nabla^{\al \bt}\nabla_{\bt 1}   V \bigg) \nn \\
&- \dth^{\al \bt} S_{\al 1} \nabla_{\bt 1} V   \bigg](\mathbf{x}, z)  \,. \label{GplusVAdS}
\end{align}
\end{subequations}

In writing eqs.~\eqref{3pointterms}, we are ignoring terms in $\mathcal{V}$ and $\big( G^+_0 \mathcal{V}\big)$ in \eqref{GplusVAdS} that do not contribute to the correlator: either due to the $\{\rho, \sg\}$-ghosts background charge saturation or because it is a total derivative (and hence a null state in the CFT). For completeness, gauge-invariance of \eqref{3pointterms} is shown in Appendix \ref{gaugeinvhybV}.

To simplify the notation, let us denote $(\mathbf{x}_i,z_i)=(\mathbf{i})$ in \eqref{3pointAdS}. The upper index in $\mathcal{V}^{(1)}$ is there to label the state, similarly for $\{\mathcal{V}^{(2)}, \mathcal{V}^{(3)} \}$. After integrating out the $\{\rho,\sg\}$-ghosts, ``integrating by parts'' to eliminate the explicit $z$ dependence and using the equation of motion $\nabla^{\al \bt} \nabla_{\al 1} \nabla_{\bt 1} V =0$, the amplitude \eqref{3pointAdS} reads\footnote{We have checked that this partial integration produces the same answer before and after integrating over the worldsheet fermions. The reason for this is that the fermionic measure is invariant under $\nabla_{\ab}$ and $\nabla_{\al 1}$, which are the zero-modes appearing in the vertices.}
\begin{align}\label{3point}
\mathcal{A}_3 & = \frac{1}{2k} \frac{1}{\sqrt{2}} \bigg[ \frac{i}{2} \bigg( \bigg\langle V^{(3)}(\mathbf{3}) \nabla^{\al \bt} V^{(2)}(\mathbf{2})\nabla_{\al 1} \nabla_{\bt 1} V^{(1)}(\mathbf{1}) \bigg\rangle \nn \\
& + \bigg\langle \nabla^{\al \bt} V^{(3)} (\mathbf{3}) \nabla_{\al 1} \nabla_{\bt 1} V^{(2)}(\mathbf{2}) V^{(1)}(\mathbf{1}) \bigg\rangle \nn \\
& + \bigg\langle \nabla_{\al 1} \nabla_{\bt 1} V^{(3)}(\mathbf{3}) V^{(2)}(\mathbf{2})\nabla^{\al \bt} V^{(1)}(\mathbf{1}) \bigg\rangle \bigg) \nn \\
& +2 \sqrt{2} \dth^{\al \bt} \bigg\langle V^{(3)}(\mathbf{3}) \nabla_{\al 1} V^{(2)}(\mathbf{2}) \nabla_{\bt 1} V^{(1)}(\mathbf{1}) \bigg\rangle\bigg]\,,
\end{align}
where we wrote it in the more symmetric form. For the latter, we used the identity
\begin{align}
& i \bigg\langle V^{(3)}(\mathbf{3}) \nabla_{\al 1}(x_1) V^{(2)}(\mathbf{2}) \Big( \nabla_{\bt 1} \nabla^{\al \bt} V^{(1)} \Big)(\mathbf{1}) \bigg\rangle \nn \\
& \qquad = \frac{i}{2} \bigg\langle \nabla^{\al \bt} V^{(3)}(\mathbf{3}) \nabla_{\al 1} \nabla_{\bt 1} V^{(2)}(\mathbf{2}) V^{(1)}(\mathbf{1}) \bigg\rangle \nn \\
& \qquad +  \frac{i}{2} \bigg\langle \nabla_{\al 1} \nabla_{\bt 1} V^{(3)}(\mathbf{3}) V^{(2)}(\mathbf{2}) \nabla^{\al \bt} V^{(1)}(\mathbf{1}) \bigg\rangle \,.
\end{align}
The last term of eq.~\eqref{3point} is not present in the flat space calculation and, therefore, it corresponds to a curvature correction.

We should underscore the fact that only the holomorphic part of the scattering amplitude $\mathcal{A}_3$ is being written. As in any closed string calculation where holomorphic/anti-holomorphic factorization takes place \cite{Schwarz:1982jn}, one needs to multiply eq.~\eqref{3point} with the corresponding right-moving contribution to get the complete answer. Strictly speaking, this means that the amplitude \eqref{3point} is $\rm PSU(1,1|2)_{\rm L} \times PSU(1,1|2)_{\rm R}$-covariant. In particular,  we remarked in our discussion of the hybrid formalism in Section \ref{sechybridinAdS3} that the $\rm PSU(1,1|2)_{\rm L}$ currents are purely holomorphic and the $\rm PSU(1,1|2)_{\rm R}$ currents purely anti-holomorphic.

\subsection{Integrating out the fermions}

Eq.~\eqref{3point} gives a $\rm PSU(1,1|2)$-covariant expression for the three-point amplitude of half-BPS states in $\rm AdS_3 \times S^3$. Let us now illustrate how the integration over the curved fermionic worldsheet variables \eqref{thetaintAdS} can be implemented with an example. 

For simplicity, we will take $V^{ (i)} = \frac{i}{2} (\ta \sg^{\ab} \ta) a_{\ab \, i}$, so that only states from the NS-sector are being considered. After integrating out the $\ta$'s using the prescription \eqref{thetaintAdS}, the amplitude \eqref{3point} for the NS states becomes
\begin{align}\label{3pointAdS32}
\mathcal{A}^{\rm NS}_{3}& = - \frac{1}{2k} \frac{1}{\sqrt{2}} {E_{ \ab}}^{ \db}(\mathbf{x}_1) {E_{\bb}}^{  \eb}(\mathbf{x}_2) {E_{\cb}}^{  \fb}(\mathbf{x}_3)  \bigg[ \eta_{\db \eb} a_3^{\cb}(\mathbf{3}) a_2^{\bb}(\mathbf{2}) \Big(D_{\fb}(-\mathbf{x}_1) a_1^{\ab}\Big)(\mathbf{1})  \nn \\
& + \eta_{\eb \fb}  a_3^{\cb}(\mathbf{3}) \Big(D_{\db}(-\mathbf{x}_2) a_2^{\bb}\Big)(\mathbf{2}) a^{\ab}_1(\mathbf{1}) + \eta_{\db \fb}  \Big( D_{\eb}(-\mathbf{x}_3) a_3^{\cb} \Big)(\mathbf{3}) a_2^{\bb}(\mathbf{2}) a_1^{\ab}(\mathbf{1}) \nn\\ 
& -\frac{1}{2} f_{\db \eb \fb}  a_3^{\cb}(\mathbf{3}) a_2^{\bb}(\mathbf{2}) a_1^{\ab}(\mathbf{1})  \bigg] \,,
\end{align}
which one can write in the more compact form as
\begin{align}\label{3pointAdS3}
\mathcal{A}_3^{\rm NS} & = -\frac{1}{2k} \frac{1}{\sqrt{2}} \bigg[ E_{\ab \bb}(\mathbf{x}_{12}) a_3^{\cb}(\mathbf{3}) a_2^{\bb}(\mathbf{2}) \Big( D_{\cb}(\mathbf{x}_{31}) a_1^{\ab} \Big) (\mathbf{1}) \nn \\
& + E_{\bb \cb}(\mathbf{x}_{23}) a_3^{\cb}(\mathbf{3}) \Big( D_{\ab} (\mathbf{x}_{12}) a_2^{\bb} \Big) (\mathbf{2}) a_1^{\ab}(\mathbf{1}) \nn \\
&+ E_{\ab \cb} (\mathbf{x}_{13}) \Big( D_{\bb} (\mathbf{x}_{23}) a_3^{\cb} \Big) (\mathbf{3}) a_2^{\bb}(\mathbf{2}) a_1^{\ab}(\mathbf{1}) \nn \\
& -\frac{1}{2} f_{\ab \bb \cb} (\mathbf{x}_1 , \mathbf{x}_2, \mathbf{x}_3)a_3^{\cb}(\mathbf{3}) a_2^{\bb}(\mathbf{2}) a_1^{\ab}(\mathbf{1})  \bigg]\,,
\end{align}
where $D_{\ab} (\mathbf{x}_{12})  = {E_{\ab}}^{\bb}(\mathbf{x}_{12}) D_{\bb}$ and $f_{\ab \bb \cb}(\mathbf{x}_1, \mathbf{x}_2, \mathbf{x}_3)$ are defined in eqs.~\eqref{covderdefn} and \eqref{structofx}, respectively. We are also using that
\begin{align}
{E_{\ab}}^{  \cb}(\mathbf{x}_i) {E_{ \bb}}^{ \db}(\mathbf{x}_j) \eta_{\cb \db} & = E_{\ab \bb}(\mathbf{x}_{ij})\,.
\end{align}

The design of the amplitude \eqref{3pointAdS3} begs for an interpretation. As we have alluded to below eq.~\eqref{mobiusgen}, the spacetime vielbein field ${E_{\ab}}^{\bb}(\mathbf{x})$ encodes that a conformal transformation on the $\rm AdS_3$ boundary corresponds to a rotation in the $\rm AdS_3$ bulk. In particular, this can be seen by the observation that the object ${E_a}^{+} \pd_+$ generates infinitesimal Möbius transformations along $\rm \pd AdS_3$.

Moreover, from eq.~\eqref{3pointAdS3}, one can explicitly deduce that the consequence of integrating out the fermionic worldsheet fields in the correlator was the appearance of the vielbein field ${E_{\ab}}^{\bb}(\mathbf{x})$. In other worlds, the tangent space vector indices were ``rotated'' by the matrix ${E_{\ab}}^{\bb}(\mathbf{x})$. This rotation also affected the indices of the structure constants $f_{\ab \bb \cb}$ which became $f_{\ab \bb \cb}(\mathbf{x}_1, \mathbf{x}_2, \mathbf{x}_3)$ of \eqref{structofx2}.

Let us point out that what is left in \eqref{3pointAdS3} is the kinematic factor of the three-point amplitude written in terms of the component fields from the NS-sector. Using the vertex operator \eqref{vertexAdS33}, one can similarly write the kinematic factor involving the states from the R-sector. 

In addition, by conformal invariance in the worldsheet and target-space, the $z$ and $\mathbf{x}$ dependence of the amplitude is completely fixed \cite[eq.~(2.13)]{Maldacena:2001km}. More precisely, the amplitude is independent of $z$, and the $\mathbf{x}$ dependence is determined by the $\rm SL(2, \mathbb{R})$ spin $j_i$ of the insertions in \eqref{3pointAdS}. So that it takes the general form \cite{Maldacena:2001km}
\begin{align}
\mathcal{A}_3 & \sim \mathbf{x}_{12}^{j_3 - j_1 -j_2} \mathbf{x}_{13}^{j_2 - j_1 -j_3} \mathbf{x}_{23}^{j_1 - j_2 -j_3}\,.
\end{align}
In Appendix \ref{MOvertices}, we give a brief explanation on how the $\rm SL(2, \mathbb{R})$ spin $j_i$ for the fermionic coordinates and component fields can be derived from our worldsheet variables. In particular, note that the variables $\ta^{\al}$ in the vertex carry a non-zero charge, see eqs.~\eqref{sl2su2}.

\section{Comparison with RNS} \label{secRNSdescription}

Up to now, in the calculations displayed throughout this work, we have used the hybrid description written in terms of the supergroup variable $g$ (or the $\rm PSU(1,1|2)$ currents) as in the worldsheet action \eqref{NSNSaction}. Even the definition of the fermionic zero-mode variable $\ta^{\al}$ in Section \ref{AdS3superfield} could be motivated in this formulation, which is the best suited for the study of the superstring in $\rm AdS_3 \times S^3$ since it generalizes to the case where a non-zero amount of R-R self-dual three-form flux is turned on \cite{Berkovits:1999im}.

That being the case, the hybrid formalism in $\rm AdS_3 \times S^3$ with pure NS-NS three-form flux can also be written in terms of bosonic currents $\mathcal{J}_{\ab}$ and free fermions $\{p_{\al},\ta^{\al}\}$. As was mentioned above eqs.~\eqref{RNScurrentalgebra}, this hinges on the fact that the matter part of the RNS formalism in the pure NS-NS $\rm AdS_3 \times S^3$ target-space is given in terms of the bosonic currents $\mathcal{J}_{\ab}$ and the six free fermions $\psi_{\ab}$ \cite{Giveon:1998ns} \cite{Ferreira:2017pgt} \cite{Sriprachyakul:2024gyl}. Therefore, the free RNS fermions $\psi_{\ab}$ plus $\{\bt, \g\}$-ghosts are related to the free fermions $\{p_{\al}, \ta^{\al}\}$ in \eqref{freefermions}.

In this section, we will further explore this correspondence between hybrid and RNS variables to compare the vertex operators and amplitude computation for the NS-sector states in Section \ref{secthreepointhyb} with the analogous calculation in the RNS formalism. Achieving the same result with a different method should give further support to our construction.

\subsection{From hybrid to RNS variables}

Let us identify, in RNS language, the contributions to the three-point amplitude \eqref{3pointAdS} for the NS-sector states. In terms of the RNS variables, the hybrid formalism worldsheet fields can be expressed as \cite{Berkovits:1994vy} \cite{Berkovits:1999im}
\begin{subequations}\label{fieldredeftoRNS}
\begin{align}
S_{\al 1}& = e^{-\frac{\phi}{2}}e^{ - \frac{i}{2} H^{\rm RNS}_C} S_{\al } \,,& \ta^{\al } & = S^{\al } e^{\frac{i}{2} H^{\rm RNS}_C} e^{\frac{\phi}{2}}  \,, \\
 e^{\rho} & = e^{-2 \phi + i \chi - i H^{\rm RNS}_C } \,,& J^{++}_C & = e^{-2 i \chi + 2 \phi + i H_C^{\rm RNS}} \,,
\end{align}
\end{subequations}
where $S_{\al}$ is the spin-field for the six-dimensional part and the boson $H_C^{\rm RNS}$ defines the spin-field for the compactified directions.\footnote{We apologyze for using the letter $S$ both for the RNS spin-field $S_{\al}$ and for the $\rm PSU(1,1|2)$ fermionic currents $S_{\al j}$. Since the currents also carry an $\rm SU(2)$ index, this notation is unambiguous.} 

The field $e^{i \sg} $ in the six-dimensional hybrid formalism is the $c$-ghost of the RNS description in bosonized form and $e^{-i \sg}$ the $b$-ghost. Similarly, the chiral bosons $\{\phi, \chi\}$ come from the superconformal ghosts 
\begin{align}
\bt &= e^{-\phi} \pd \xi = e^{-\phi} \pd e^{i \chi} \,,&\g &=\eta e^{\phi}= e^{-i\chi} e^{\phi}.
\end{align}
Note that the RNS variables obey the usual OPEs
\begin{subequations}
\begin{align}
H_C^{\rm RNS} (y) H_C^{\rm RNS} (z) & \sim -2 \log (y-z) \,,& \sg(y) \sg(z) & \sim - \log(y-z)\,, \\
\phi(y) \phi (z) & \sim - \log(y-z) \,,& \chi(y) \chi(z) & \sim - \log(y-z)\,.
\end{align}
\end{subequations}

Consequently, in terms of the RNS description, we have that
\begin{subequations}
\begin{align}
\mathcal{V} & = \xi V_{\rm hyb}^{-1}\,, \\
\widetilde{G}^+_0 \mathcal{V} & = V^{-1}_{\rm hyb} \,, \\
G^+_0 \mathcal{V} & = V^0_{\rm hyb} \,,
\end{align}
\end{subequations}
where
\begin{subequations}\label{RNSvertexfromhyb}
\begin{align}
V^{-1}_{\rm hyb} & = c \psi_{\ab} e^{-\phi} a^{\ab} \,, \\
V^0_{\rm hyb} & = - \frac{1}{2k} \frac{1}{\sqrt{2}} c \Big( K_{\ab} a^{\ab} + \psi^{\ab} \psi^{\bb} \nabla_{\ab} a_{\bb} \Big)\,, \label{V0hyb}
\end{align}
\end{subequations}
with $V^{-1}_{\rm hyb}$ and $V^0_{\rm hyb}$ being vertex operators in the $-1$ and zero picture for the NS-sector massless states, respectively. The ${\rm SU(1,1)}_{k+2} \times {\rm SU(2)}_{k-2} $ current $\mathcal{J}_{\ab}$ decouples from the fermions and is defined in eqs.~\eqref{RNScurrents1} and \eqref{RNScurrents2}. We emphasize that no excitations in the compactified directions are being considered.

To get to eqs.~\eqref{RNSvertexfromhyb}, we used the following identifications between the RNS and hybrid fermionic fields in the vertex operators
\begin{subequations}
\begin{align}
{f_{\ab \, \al 1}}^{\bt 1} S_{\bt1} \ta^{\al} & = \frac{1}{2} f_{\ab \bb \cb} \psi^{\cb} \psi^{\bb} \,, \\ 
(\sg^{\bb \cb})^{\bt}_{\ \al} S_{\bt 1} \ta^{\al} & = - i \psi^{\bb} \psi^{\cb} \,, \\
K_{\ab} & = \mathcal{J}_{\ab} + \frac{1}{2} f_{\ab \bb \cb} \psi^{\cb} \psi^{\bb}\,. \label{RNSboscurrent1}
\end{align}
\end{subequations}

\subsection{RNS formalism in $\rm AdS_3 \times S^3$}

Of course, one can arrive at the vertex operators \eqref{RNSvertexfromhyb} directly from the RNS description of $\rm AdS_3 \times S^3$, which is given by a bosonic ${\rm SU(1,1)}_{k+2} \times {\rm SU(2)}_{k-2}$ current algebra plus six free fermions $\psi_{\ab}$. Needless to say, one should consider the GSO projected theory in order to eliminate the tachyons in RNS \cite{Friedan:1985ge} \cite{Giveon:1998ns}. This comes in contrast with the hybrid description, in which the physical states are automatically GSO projected \cite{Berkovits:1996bf}.

In terms of the RNS variables, the bosonic currents of ${\rm PSU(1,1|2)}_k$ are the same as in \eqref{RNSboscurrent1} and read
\begin{align}\label{boscurrentRNS}
K_{\ab} & = \mathcal{J}_{\ab} + \frac{1}{2} f_{\ab \bb \cb} \psi^{\cb} \psi^{\bb} \,,
\end{align}
where the currents $\mathcal{J}_{\ab}$ are defined in eqs.~\eqref{RNScurrents1} and \eqref{RNScurrents2} and $\psi_{\ab}$ are the six free worldsheet fermions of the RNS formalism satisfying the usual OPE relation
\begin{align}
\psi_{\ab} (y) \psi_{\bb}(z) & \sim (y-z)^{-1} \eta_{\ab \bb}\,.
\end{align}
Under the $\rm PSU(1,1|2)$ bosonic currents, the fermions transform in the adjoint representation
\begin{align}
K_{\ab} (y) \psi_{\bb}(z) & \sim (y-z)^{-1} {f_{\ab \bb}}^{\cb}\psi_{\cb} \,.
\end{align}
Recall that the structure constants are defined in eqs.~\eqref{PSUstructureconst}.

The currents $\mathcal{J}_{\ab}$ have no poles with the fermions $\psi_{\ab}$ and the $\mathcal{N}=1$ supercurrent of the RNS formalism for the $\rm AdS_3 \times S^3$ part is
\begin{align}
G_6 =  \frac{i}{\sqrt{2k}} \bigg( \mathcal{J}^{\ab} \psi_{\ab} + \frac{1}{6} f_{\ab \bb \cb} \psi^{\cb} \psi^{\bb} \psi^{\ab} \bigg) \,,
\end{align}
that, together with the stress-tensor,
\begin{align}
T_6 & = - \frac{1}{4k} \mathcal{J}_{\ab} \mathcal{J}_{\bb} \eta^{\ab \bb} - \frac{1}{2} \psi_{\ab} \pd \psi_{\bb} \eta^{\ab \bb}\,,
\end{align}
generate a $c=9$ $\mathcal{N}=1$ SCA. The four bosons and four fermions for the compactification directions generate a $c=6$ $\mathcal{N}=1$ SCA whose supercurrent and stress-tensor we denote by $G_C^{\rm RNS}$ and $T_C^{\rm RNS}$, respectively. In total, one has the usual matter $c=15$ $\mathcal{N}=1$ SCA of the RNS description
\begin{subequations}
\begin{align}
T_{\rm m}(y)T_{\rm m}(z) & \sim \frac{\frac{c}{2}}{(y-z)^4} + \frac{2T_{\rm m}(z)}{(y-z)^2} + \frac{\pd T_{\rm m}(z)}{(y-z)} \,, \\
T_{\rm m}(y)G_{\rm m}(z) & \sim \frac{\frac{3}{2}G_{\rm m}(z)}{(y-z)^2}+ \frac{\pd G_{\rm m}(z)}{(y-z)} \,, \\
G_{\rm m}(y)G_{\rm m}(z) & \sim \frac{\frac{2}{3}c}{(y-z)^3} + \frac{2T_{\rm m}(z)}{(y-z)} \,,
\end{align}
\end{subequations}
with generators $\{G_{\rm m}=G_6 + G_C^{\rm RNS}, T_{\rm m}=T_6 + T_C^{\rm RNS}\}$.

The NS-sector massless unintegrated vertex operators in the $-1$ and zero picture are
\begin{subequations}\label{RNSVpics}
\begin{align}
V^{-1}_{\rm RNS} & = c \psi_{\ab} e^{-\phi} a^{\ab} \,, \\ 
V^{0}_{\rm RNS} & = -\frac{1}{2k} \frac{1}{\sqrt{2}} c \Big( \mathcal{J}_{\ab} a^{\ab} + \frac{1}{2}f_{\ab \bb \cb} \psi^{\cb} \psi^{\bb} a^{\ab} + \psi^{\ab} \psi^{\bb} (\mathcal{J}_{\ab})_0 a_{\bb}  \Big)\nn \\
& = -\frac{1}{2k} \frac{1}{\sqrt{2}} c \Big( K_{\ab} a^{\ab} + \psi^{\ab} \psi^{\bb} (\mathcal{J}_{\ab})_0 a_{\bb}  \Big)\label{V0RNS}\,,
\end{align}
\end{subequations}
where $(\mathcal{J}_{\ab})_0$ is the zero-mode of the current $\mathcal{J}_{\ab}$ and $K_{\ab}$ is defined in eq.~\eqref{boscurrentRNS}. Up to a constant $V^0_{\rm RNS} = Z V^{-1}_{\rm RNS}$, where
\begin{align} 
Z & = 2 Q_{\rm RNS}  e^{i \chi} \nn \\
& = G_{\rm m} e^{\phi}  + b \pd e^{- i \chi} e^{2 \phi} + \frac{1}{2} \pd \big( b e^{-i \chi} e^{ 2 \phi} \big) + 2 c \pd e^{i \chi}\,,
\end{align}
is the picture-changing operator \cite{Friedan:1985ge}. The BRST operator in the RNS formalism is\footnote{The option for the total derivative added in the BRST current $j_{\rm BRST}$ is chosen such that the double pole between $j_{\rm BRST}$ and $b$ is given by the ghost- minus the picture-current. For that reason, $j_{\rm BRST}$ gets mapped to the $\mathcal{N}=2$ superconformal generator $G^+$ \eqref{GplusAdS3}.}
\begin{align}
Q_{\rm RNS} & = \oint j_{\rm BRST} \nn \\
& = \oint \bigg( c \big( T_{\rm m} +  T_{\phi, \chi} \big)  + bc\pd c - \frac{1}{2} e^{-i \chi + \phi} G_{\rm m} + \frac{1}{4} b e^{-2i \chi + 2 \phi} + \pd^2 c  +   \pd \big( \pd(i\chi) c \big) \bigg) \,.
\end{align}

Since the zero-mode o $\mathcal{J}_{\ab}$ acts on $a_{\ab}$ as the zero-mode of $K_{\ab}$. We can write
$V_0^{\rm RNS}$ in the form
\begin{align}
V_0^{\rm RNS} &  = - \frac{1}{2k} \frac{1}{\sqrt{2}} c \Big( K_{\ab} a^{\ab} + \psi^{\ab} \psi^{\bb} \nabla_{\ab} a_{\bb} \Big)\,, 
\end{align}
which precisely matches the vertex operator \eqref{V0hyb} found by the field redefinition from the hybrid formalism.

\subsection{Three-point amplitude in RNS variables} \label{RNSampAdS3}

We can now use the tools developed in this section to compute the three-point amplitude \eqref{3pointAdS3} for the NS-sector states inserted on the $\rm AdS_3$ boundary directly in terms of the RNS formalism prescription.

As before, for fields depending on the boundary coordinates, we have
\begin{align}
\psi_{\ab} (\mathbf{x}) & = e^{\mathbf{x} \nabla_+} \psi_{\ab} e^{-\mathbf{x} \nabla_+} \nn \\
& = {E_{\ab}}^{\bb}(\mathbf{x}) \psi_{\bb}\,,
\end{align}
where ${E_{\ab}}^{\bb}(\mathbf{x})$ is given by \eqref{vielbeinfieldAdS3}. Hence, the fundamental OPEs read
\begin{subequations}
\begin{align}
\psi_{\ab} (\mathbf{x}_i,y) \psi_{\bb} (\mathbf{x}_j,z) & \sim (y-z)^{-1} E_{\ab \bb}(\mathbf{x}_{ij}) \,, \\
K_{\ab}(\mathbf{x}_i,y) \psi_{\bb}(\mathbf{x}_j,z) & \sim (y-z)^{-1} {E_{\ab}}^{ \cb} (\mathbf{x}_i) {E_{\bb}}^{\db} (\mathbf{x}_j) {f_{\cb \db}}^{\eb} \psi_{\eb} \,.
\end{align}
\end{subequations}

Considering the vertex operators \eqref{RNSVpics}, the three-point amplitude for the NS-sector states becomes
\begin{align}
\mathcal{A}^{\rm NS , RNS}_3 & = \bigg\langle V_{-1}^{\rm RNS} (\mathbf{3}) V_{-1}^{\rm RNS} (\mathbf{2}) V_0^{\rm RNS}(\mathbf{1}) \bigg\rangle \nn \\
& = -\frac{1}{2k} \frac{1}{\sqrt{2}} \bigg[ E_{\ab \bb}(\mathbf{x}_{12}) a_3^{\cb}(\mathbf{3}) a_2^{\bb}(\mathbf{2}) \Big( D_{\cb}(\mathbf{x}_{31}) a_1^{\ab} \Big) (\mathbf{1}) \nn \\
& + E_{\bb \cb}(\mathbf{x}_{23}) a_3^{\cb}(\mathbf{3}) \Big( D_{\ab} (\mathbf{x}_{12}) a_2^{\bb} \Big) (\mathbf{2}) a_1^{\ab}(\mathbf{1}) \nn \\
&+ E_{\ab \cb} (\mathbf{x}_{13}) \Big( D_{\bb} (\mathbf{x}_{23}) a_3^{\cb} \Big) (\mathbf{3}) a_2^{\bb}(\mathbf{2}) a_1^{\ab}(\mathbf{1}) \nn \\
& -\frac{1}{2} f_{\ab \bb \cb} (\mathbf{x}_1 , \mathbf{x}_2, \mathbf{x}_3)a_3^{\cb}(\mathbf{3}) a_2^{\bb}(\mathbf{2}) a_1^{\ab}(\mathbf{1})  \bigg]\,,
\end{align}
which precisely matches \eqref{3pointAdS3}, as we wanted to show. This calculation gives further support for our construction using the supergroup variables and the fermionic zero-mode coordinates $\ta^{\al}(\mathbf{x})$ in the hybrid formalism.  

Lastly, let us mention that, under the field redefinition \eqref{fieldredeftoRNS}, the RNS tree-level zero-mode integration  gets mapped to the hybrid measure of Section \ref{treeampprescription} only if one works in the large Hilbert space, namely,
\begin{align}
\xi c \pd c \pd^2 c e^{-2\phi} & \sim e^{3\rho +3 i \sg} J^{++}_C (\ta)^4 \,.
\end{align}

\section{Conclusion} \label{secconc}

In this work, we have studied the superstring in the $\rm AdS_3 \times S^3 \times \mathcal{M}_4$ background (where $\mathcal{M}_4$ is a Calabi-Yau two-fold) with $k$ units of NS-NS self-dual three-form flux from the $\rm PSU(1,1|2)$ supergroup perspective. We used the Berkovits-Vafa-Witten hybrid formalism in $\rm AdS_3$ \cite{Berkovits:1999im} and wrote a  $\rm PSU(1,1|2)$-covariant three-point amplitude for half-BPS states inserted on the $\rm AdS_3$ boundary, whose coordinates are labelled by $\mathbf{x}$. As a corollary, we found that the kinematic factor gets dressed with the vielbein ${E_A}^B(\mathbf{x})$ after the worldsheet fermions are integrated out in the path-integral. In addition, we saw the compelling fact of the conformal group on the boundary being identified with the symmetry group of the $\rm AdS_3$ bulk by explicitly analyzing the form of ${E_A}^B(\mathbf{x})$, which naturally appears in our covariant superstring description. We ended by showing that our results agree with the RNS formalism answer.

Let us now comment on some related subjects that deserve future investigation. An immediate application of this work is the computation of the amplitude expression in components involving two R-sector states and one NS-sector state starting from the superspace form \eqref{3point}. 

Although string theory scattering amplitudes in flat space  have been extensively explored both at tree- and quantum-level \cite{Berkovits:2022ivl} \cite{Mafra:2022wml}, superstring amplitudes in curved backgrounds are still a mysterious subject. In view of the $\rm AdS/CFT$ correspondence, it is highly desirable to have superstring perturbation theory under good control in an $\rm AdS$ background.\footnote{See ref.~\cite{Alday:2023mvu} for recent progress in the $\rm AdS_5 \times S^5$ Virasoro-Shapiro amplitude.} That being said, the construction presented in this paper provides all the necessary elements for the exploration of higher-point $\rm PSU(1,1|2)$-covariant tree-level scattering amplitudes in $\rm AdS_3 \times S^3$. 

To be more specific, we gave a zero-mode prescription for the fermionic worldsheet fields \eqref{thetaintegration} with vertex operators depending on the $\rm AdS_3$ boundary coordinates $\mathbf{x}$. The remaining ingredient is the integrated vertex operator \eqref{intvertexAdS3}. For the half-BPS states \eqref{vertexAdS34}, the holomorphic part reads
\begin{align}\label{intvertexAdS32}
\int G^+_0 G^-_{-1}  \mathcal{V} & = \int \frac{1}{2k} \bigg[ \frac{1}{\sqrt{2}} \bigg( \frac{i}{2} K^{\al \bt} \nabla_{\al 1} \nabla_{\bt 1} + i S_{\al 1} \nabla^{\al \bt} \nabla_{\bt 1} \bigg) - \dth^{\al \bt} S_{\al 1} \nabla_{\bt 1} \bigg] V \,,
\end{align}
where we only wrote the terms that contribute to the tree-level amplitudes of half-BPS states in \eqref{intvertexAdS32}.

When defining the vertex operators in Section \ref{AdS3vertexsec}, we did not consider spectrally flowed states in $ {\rm AdS}_3$ \cite{Maldacena:2000hw}. It would be interesting to determine how the fermionic worldsheet coordinates $\ta^{\al}$ behave under spectral flow, so that $\rm PSU(1,1|2)$-covariant amplitudes of spectrally flowed states could be evaluated. Given that the fermionic worldsheet fields in the hybrid formalism are related to the $\psi^{\ab}$'s in RNS, the technology developed in ref.~\cite{Sriprachyakul:2024gyl} should play an important role in this matter. Since the integration over the fermions in RNS can be expressed in terms of the covering map data \cite{Sriprachyakul:2024gyl}, one could also explore the intriguing relation between the vielbein ${E_A}^B(\mathbf{x})$ and the covering map \cite{Eberhardt:2019ywk} \cite{Dei:2020zui}.

What happens when switching on R-R flux? Although the holomorphic/anti-holomorphic factorization is lost \cite{Benichou:2010rk} \cite{Eberhardt:2018exh}, the tree-level amplitude prescription to integrate over the fermions in Section \ref{treeampprescription} should not change when turning on a non-zero amount of R-R three-form flux \cite{Bobkov:2002bx}. Given that the worldsheet action in a R-R $\rm AdS_3 \times S^3$ background is known \cite{Berkovits:1999im}, one could look for reproducing the results from refs.~\cite{Cho:2018nfn} \cite{Gaberdiel:2023lco} \cite{Frolov:2023pjw} using the spacetime supersymmetric hybrid formalism.

Finally, one could ask what can we learn from this work about the correct fermionic measure in the $\rm AdS_5 \times S^5$ pure spinor formalism? In a flat background, the pure spinor measure for the fermions picks the factor with $(\ta)^5 (\tah)^5$ in scattering amplitude computations \cite{Berkovits:2000fe}.\footnote{Here, $\ta$ is a left-moving and $\tah$ a right-moving worldsheet variable. Of course, our notation is schematic in this section.} In effect, this is in agreement with the integration prescription of the $\rm U(5)$-hybrid formalism, see \cite[eq.~(6.3)]{Berkovits:2000fe}. For the six-dimensional case, as we have shown, the flat background tree-level prescription to integrate over the $\ta$'s naturally generalize to $\rm AdS_3 \times S^3$ in terms of the standard spacetime SUSY generator in the hybrid formalism $\nabla_{\al 1}$ \eqref{thetaintegration}. 

So it seems reasonable to suspect that there is a prescription in the $\rm AdS_5 \times S^5$ pure spinor formalism which picks out a factor of $(\ta)^5 (\tah)^5$. In the $\rm U(5)$ notation, by using the standard spacetime SUSY generator $\nabla_{a}$, the prescription may take the form
\begin{align}\label{U5prescription}
\ep^{abcde} \ep^{\widehat{a}\widehat{b}\widehat{c}\widehat{d}\widehat{e}} \nabla_{a}\nabla_{b} \nabla_{c} \nabla_{d}\nabla_{e}\nabla_{\widehat{a}}\nabla_{\widehat{b}}\nabla_{\widehat{c}}\nabla_{\widehat{d}}\nabla_{\widehat{e}}\,,
\end{align}
where $a,\widehat{a} = 1$ to $5$ are fundamental and anti-fundamental $\rm U(5)$ indices in \eqref{U5prescription}. It would be interesting to first study these claims in the analogue of the $\rm AdS_5 \times S^5$ pure spinor formalism in $\rm AdS_3 \times S^3$ \cite{Berkovits:1999du} \cite{Daniel:2024yfr} by building up on the results of this paper. Note also that progress towards $\rm AdS_5 \times S^5$ vertex operators was achieved in refs.~\cite{Berkovits:2019rwq} \cite{Fleury:2021ieo}.

\section*{Acknowledgements}
CAD would like to thank João Gomide for discussions and especially Nathan Berkovits, Matthias Gaberdiel, Lucas Martins, Vit Sriprachyakul and Dennis Zavaleta for discussions and comments on the draft of this paper. CAD would also like to thank FAPESP grant numbers 2022/14599-0 and 2023/00015-0 for financial support.

\appendix

\section{\boldmath Six-dimensional Pauli matrices} \label{sigmas}
The Lorentz group $\rm SO(1, 5)$ is locally isomorphic to \text{SU}(4) and, under this identification, spinors of $\rm SO(1, 5)$ transform as $\boldsymbol{4}$'s or $\boldsymbol{4}^\prime$'s of \text{SU}(4). By definition, Weyl spinors transform as a $\boldsymbol{4}$ and are denoted by an upper lower case greek index ranging from 1 to 4. Anti-Weyl spinors transform as a $\boldsymbol{4}^\prime$ and are denoted by a down lower case greek index ranging from 1 to 4. All other representations of $\rm SO(1, 5)$ can be built from tensor products of $\boldsymbol{4}$'s and $\boldsymbol{4}^\prime$'s. The following tensor products are of particular importance
\begin{subequations} \label{tensorsg}
\begin{align}
\boldsymbol{4} \otimes \boldsymbol{4} & \simeq \boldsymbol{6} \oplus \boldsymbol{10}_- \,,  \\
\boldsymbol{4}^\prime \otimes \boldsymbol{4}^\prime & \simeq \boldsymbol{6} \oplus \boldsymbol{10}_+ \,, \\
\boldsymbol{4} \otimes \boldsymbol{4}^\prime& \simeq \boldsymbol{1} \oplus \boldsymbol{15}\,, 
\end{align}
\end{subequations}
where $\boldsymbol{1}$ denotes the singlet representation, $\boldsymbol{6}$ is antisymmetric in the spinor indices and denotes the vector representation, $\boldsymbol{10}_-$ and $\boldsymbol{10}_+$ are symmetric and correspond to anti-self-dual and self-dual three-forms, respectively,\footnote{Note that $(\sg_{012})^{\al \bt} =- (\sg^{345})^{\al \bt}$ and $(\sg_{012})_{\al \bt} = (\sg^{345})_{\al \bt}$ (see eqs.~\eqref{2and3forms}).} and the traceless representation $\boldsymbol{15}$ is a two-form.

The \text{SO}(1,5) Pauli matrices are defined as
\begin{align}\label{6dPaulisapp}
\sigma_{\al \bt}^{0}& = \frac{1}{\sqrt{2}}
\begin{pmatrix}
\boldsymbol{\sg}^2 & 0\\
0 & \boldsymbol{\sg}^2
\end{pmatrix}\,,
& \sg^1_{\al \bt} & =\frac{1}{\sqrt{2}}
\begin{pmatrix}
0& \boldsymbol{\sg}^1 \\
- \boldsymbol{\sg}^1 & 0
\end{pmatrix}\,, \nn \\
\sigma_{\al \bt}^{2}& =\frac{1}{\sqrt{2}}
\begin{pmatrix}
0 & -\boldsymbol{\sg}^2 \\
-\boldsymbol{\sg}^2 & 0
\end{pmatrix}\,,
& \sg^3_{\al \bt} & =\frac{1}{\sqrt{2}}
\begin{pmatrix}
0 & \boldsymbol{\sg}^3 \\
-\boldsymbol{\sg}^3 & 0
\end{pmatrix}\,, \\
\sigma_{\al \bt}^{4}& =\frac{1}{\sqrt{2}}
\begin{pmatrix}
0 & -i \mathbb{1} \\
i \mathbb{1} & 0
\end{pmatrix}\,,
& \sg^5_{\al \bt} & =\frac{1}{\sqrt{2}}
\begin{pmatrix}
\boldsymbol{\sg}^2 & 0 \\
0 & -\boldsymbol{\sg}^2
\end{pmatrix}\,, \nn
\end{align}
where the $\boldsymbol{\sg}$-matrices are the usual $\text{SU}(2)$ Pauli matrices
\begin{align}
\boldsymbol{\sg}^1&=
\begin{pmatrix}
0& 1 \\
1 & 0
\end{pmatrix}\,, & 
\boldsymbol{\sg}^2 &=
\begin{pmatrix}
0& -i \\
i & 0
\end{pmatrix}\,, &
\boldsymbol{\sg}^3 &=
\begin{pmatrix}
1& 0 \\
0 & -1
\end{pmatrix}\,.
\end{align}
The $\sg$-matrices are antisymmetric and satisfy the algebra
\begin{equation}
\sg^{\ab \al \bt} \sg^{\bb}_{\al \g} + \sg^{\bb \al \bt}\sg^{\ab}_{\al \g}= \eta^{\ab \bb} \dt^\bt_\g \,,
\end{equation}
where $\eta^{\ab \bb}= \text{diag}(-,+,+,+,+,+)$, $\ab=\{0$ to $5\}$, is the six-dimensional Minkowski metric and we define
\begin{equation}
\sg^{\ab \al \bt} = \frac{1}{2} \ep^{\al \bt \g \dt} \sg^{\ab}_{\g \dt}\,,
\end{equation}
which are given by
\begin{align}
\sigma^{0 \al \bt}& = \frac{1}{\sqrt{2}}
\begin{pmatrix}
\boldsymbol{\sg}^2 & 0\\
0 & \boldsymbol{\sg}^2
\end{pmatrix}\,,
& \sg^{1 \al \bt} & =\frac{1}{\sqrt{2}}
\begin{pmatrix}
0& \boldsymbol{\sg}^1 \\
- \boldsymbol{\sg}^1 & 0
\end{pmatrix}\,, \nn \\
\sigma^{2 \al \bt}& =\frac{1}{\sqrt{2}}
\begin{pmatrix}
0 & \boldsymbol{\sg}^2 \\
\boldsymbol{\sg}^2 & 0
\end{pmatrix}\,,
& \sg^{3 \al \bt} & =\frac{1}{\sqrt{2}}
\begin{pmatrix}
0 & \boldsymbol{\sg}^3 \\
- \boldsymbol{\sg}^3 & 0
\end{pmatrix}\,, \\
\sigma^{4 \al \bt}& =\frac{1}{\sqrt{2}}
\begin{pmatrix}
0 & i \mathbb{1} \\
-i \mathbb{1} & 0
\end{pmatrix}\,,
& \sg^{5 \al \bt} & =\frac{1}{\sqrt{2}}
\begin{pmatrix}
- \boldsymbol{\sg}^2 & 0 \\
0 & \boldsymbol{\sg}^2
\end{pmatrix}\,. \nn
\end{align}
It is convenient to introduce the unitary matrix $B$, also known as an intertwiner,
\begin{align}
B_{\al}^{\ \bt}&=-(B^*)_{\al}^{\ \bt}=
\begin{pmatrix}
\boldsymbol{\sg}^2 & 0\\
0 & \boldsymbol{\sg}^2
\end{pmatrix}\,, & (B^*)_{\al}^{\ \bt} B_{\bt}^{\ \g}& = - \dt_{\al}^{\g}\,, \label{Bmatrix}
\end{align} 
so that
\begin{equation}
(\sg^{\ab}_{\al \bt})^{*} = (B)_{\al}^{\ \g} (B)_{\bt}^{\ \dt} \sg^{\ab}_{\g \dt}\,.
\end{equation}

We also define
\begin{subequations}\label{2and3forms}
\begin{align}
(\sg^{\ab \bb})_{\al}^{\ \bt} &= \frac{i}{2} (\sg^{[\ab} \sg^{\bb] })_{\al}^{\ \bt}\,, \\
(\sg^{\ab \bb \cb})^{\al \bt} & = \frac{i}{3!}(\sg^{[\ab}\sg^{\bb}\sg^{\cb]})^{\al \bt}\,,
\label{lorentzgennew}
\end{align}
\end{subequations}
where we anti-symmetrize/symmetrize without dividing by the number of terms. The Lorentz generators satisfy the commutators
\begin{subequations}
\begin{align}
[\sg_{\ab} , \sg_{\bb \cb}] & = -i \eta_{\ab [ \bb} \sg_{\cb]}\,, \\
[\sg_{\ab \bb}, \sg_{\cb \db}] & = \frac{i}{2} \Big( \eta_{ \cb [ \ab} \dt^{[\eb}_{\bb]} \dt^{\fb]}_{\db} + \eta_{\db[\bb} \dt^{[\eb}_{\ab ]} \dt^{\fb]}_{\cb} \Big) \sg_{\eb \fb} \nn \\
& = i \eta_{\cb [\ab} \sg_{\bb]\db} - i \eta_{\db[\ab} \sg_{\bb]\db}\,. 
\end{align}
\end{subequations}

Some useful identities are
\begin{subequations}
\begin{align}
\sigma^{\ab}_{\al \bt} \sg^{\bb}_{\g \dt} \eta_{\ab \bb} &= \ep_{\al \bt \g \dt}\,, \\
\sg^{\ab \al \bt} \sg^{\bb}_{\al \g} \eta_{\ab \bb} &= 3 \dt^\bt_\g\,, \\
\sg^{\ab \al \bt} \sg^{\bb}_{\al \bt} &= 2 \eta^ {\ab \bb}\,, \\
\sg^{\ab \al \bt} \sg^{\bb}_{\g \dt} \eta_{\ab \bb} & = \dt^{\al}_{\g} \dt^{\bt}_{\dt} - \dt^{\bt}_{\g} \dt^{\al}_{\dt}\,, \\
\ep^{\al \bt \rho \sg} \ep_{ \g \dt \rho \sg} & = 2 (\dt^{\al}_{\g} \dt^{\bt}_{\dt} - \dt^{\bt}_{\g} \dt^{\al}_{\dt})\,, \\
\ep^{\al \bt \g \dt} \sg_{\ab \dt \sg} & =  -\frac{1}{2} \dt_{\sg}^{[\al} \sg_{\ab}^{\bt \g ]} \,,  \\
(\sg^{\ab \bb})^{\al}_{\ \bt} (\sg^{\cb \db})^{\bt}_{ \ \al} & =\eta^{\ab [\cb} \eta^{ \db]\bb}\,, \\
\eta_{\ab \cb} \eta_{\bb \db} (\sg^{\ab \bb})_{\bt}^{\ \al} (\sg^{\cb \db})_{\g}^{\ \dt} & = - \frac{1}{2} \dt^{\al}_{\bt} \dt^{\dt}_{\g} + 2 \dt^{\al}_{\g} \dt^{\dt}_{\bt} \,, \\
(\sg_{\ab} \sg_{\bb} \sg_{\cb} \sg_{\db} )_{\al}^{\ \al} & = \eta_{\ab \bb} \eta_{\cb \db} + \eta_{\ab \db} \eta_{\bb \cb} - \eta_{\ab \cb} \eta_{\bb \db} \,, \\
(\sg_{\ab \bb \cb})_{\g \dt} \sg^{\cb}_{\al \bt} & =- \frac{i}{2} \sg_{[\ab | \al (\g |} \sg_{| \bb] | \dt) \bt}\,, \\
(\sg^{\ab \bb})^{\g}_{\ \dt} (\sg_{\ab \bb \cb})_{\al \bt} & = - \sg_{\cb \dt ( \al} \dt^{\g}_{\bt)}\,,
\end{align}
\end{subequations}
where $\ep_{1234}=1$.

Note further that the antisymmetric tensors $\ep_{\al \bt \g \dt}$ and $\ep_{jk}$ satisfy the Schouten identities
\begin{subequations}
\begin{align}
\dt^{\sg}_{[\al} \ep_{\bt \g \dt \rho]} & = 0\,, \label{schout} \\
\ep_{j[k} \ep_{lm]} & = 0\,, \label{schout2}
\end{align}
\end{subequations}
and, in addition, we have
\begin{align} \label{idep}
\ep^{jk}\ep_{lm} & = - ( \dt^j_l \dt^k_m - \dt^k_l \dt^j_m)\,.
\end{align}

Some additional trace identities are
\begin{align}
& (\sg^{\ab}\sg^{\bb}\sg^{\cb}\sg^{\db}\sg^{\eb}\sg^{\fb})^{\al}_{\ \al}  \nn \\
& \qquad = - \frac{1}{2} \ep^{\ab \bb \cb \db \eb \fb} + \frac{1}{2} \eta^{\ab \db} \eta^{\eb[\bb}\eta^{\cb]\fb} - \frac{1}{2} \eta^{\bb \db} \eta^{\eb[\ab} \eta^{\cb]\fb} + \frac{1}{2} \eta^{\cb \db} \eta^{\eb[\ab} \eta^{\bb]\fb} \nn \\
& \qquad + \frac{1}{2} \eta^{\bb \cb} \big( -\eta^{\ab \db} \eta^{\eb \fb} + \eta^{\ab \eb} \eta^{\db \fb} - \eta^{\ab \fb} \eta^{\db \eb} \big) + \frac{1}{2} \eta^{\ab \cb} \big( \eta^{\bb \db} \eta^{\eb \fb}  \nn \\
& \qquad - \eta^{\bb \eb} \eta^{\db \fb} + \eta^{\bb \fb} \eta^{\db \eb} \big) + \frac{1}{2} \eta^{\ab \bb} \big( - \eta^{\cb \db} \eta^{\eb \fb} + \eta^{\cb \eb} \eta^{\db \fb} - \eta^{\cb \fb} \eta^{\db \eb} \big) \,,
\end{align}
and
\begin{align}
(\sg_{\ab \bb \cb})^{\al \bt} (\sg_{\db \eb \fb})_{\bt \al} & = \frac{1}{2} \ep_{\ab \bb \cb \db \eb \fb} - \frac{1}{2} \eta_{[\ab|\db} \eta_{|\bb|\eb} \eta_{|\cb]\fb}\,,
\end{align}
where $\ep_{012345}=-\ep^{012345}=1$.

\section{The supercurrent $G^+$} \label{Gplusidapp}

Let us prove eq.~\eqref{Gplusid}, namely,
\begin{align}
G^+ & = - \frac{1}{4k}  \frac{1}{24} \ep^{\al \bt \g \dt} Q_{\al 2} Q_{\bt 2} Q_{\g 2} Q_{\dt 2} e^{2 \rho + 3 i \sg} + G^+_C\,,
\end{align}
where $Q_{\al 2}  = \oint \big( S_{\al 1} e^{-\rho -i \sg} + S_{\al 2} \big) $. Using the current algebra \eqref{currentalg2}, we start by noting that
\begin{subequations}
\begin{align}
Q_{\dt 2} e^{2 \rho +3 i \sg} & = S_{\dt 1} e^{\rho + 2i \sg} \,, \\
Q_{\g 2} Q_{\dt 2} e^{2 \rho + i \sg} &= -S_{\g 1} S_{\dt 1} e^{i \sg} - i\sqrt{2} K_{\g \dt} e^{\rho + 2 i \sg} \,, \\
Q_{\bt 2} Q_{\g 2} Q_{\dt 2} e^{2 \rho + i \sg} & = - S_{\bt 1} S_{\g 1} S_{\dt 1} e^{-\rho} + i \sqrt{2} \big(K_{\bt \g} S_{\dt 1} - S_{\g 1} K_{\bt \dt} + K_{\g \dt} S_{\bt 1} \big) e^{i \sg}  \nn \\
& - 2 \ep_{\bt \g \dt \rho} \dth^{\rho \sg} \big[ \big( S_{\sg 1} e^{-\rho -i \sg}, e^{\rho + 2i \sg} \big) + S_{\sg 2} e^{\rho + 2 i \sg} \big]\,.
\end{align}
\end{subequations}
Therefore,
\begin{align}
&- \frac{1}{4k} \frac{1}{24} \ep^{\al \bt \g \dt} Q_{\al 2} Q_{\bt 2} Q_{\g 2} Q_{\dt 2} e^{2 \rho + 3 i \sg} \nn\\
& \qquad  = - \frac{1}{4k} (S_1)^4 e^{-2 \rho -i \sg} -  \frac{1}{2k}\bigg( \frac{i}{2\sqrt{2}} S_{\al 1} S_{\bt 1}K^{\al \bt}  +\dth^{\al \bt} S_{\al 1} \pd S_{\bt 1} \bigg) e^{-\rho} \nn \\
& \qquad + T_{\rm PSU} e^{i \sg} +  \big(\pd e^{-\rho -i \sg}, e^{\rho + 2i \sg} \big) \,,
\end{align}
implying we can write eq.~\eqref{GplusAdS3} as \eqref{Gplusid}, as we wanted to show.

\section{\boldmath Twisted small $\mathcal{N}=4$ SCA}\label{N=4algApp}

The twisted $\mathcal{N}=2$ superconformal algebra with central charge $c$ satisfied by the generators $\{J,G^+,G^-,T\}$ is given by
\begin{subequations} 
\begin{align}
T(y)T(z) & \sim \frac{2 T(z)}{(y-z)^2} +\frac{ \pd T(z)}{(y-z)} \,, \\
G^+(y)G^-(z) & \sim \frac{\frac{c}{3}}{(y-z)^3} + \frac{J(z)}{(y-z)^2} + \frac{T(z)}{(y-z)}\,, \\
T(y)G^{+}(z) & \sim \frac{G^{+}(z)}{(y-z)^2} + \frac{\pd G^{+}(z)}{(y-z)}\,, \\
T(y)G^{-}(z) & \sim \frac{2G^{-}(z)}{(y-z)^2} + \frac{\pd G^{-}(z)}{(y-z)}\,, \\
T(y)J(z) & \sim - \frac{\frac{c}{3}}{(y-z)^{3}} + \frac{J(z)}{(y-z)^2} + \frac{ \pd J(z)}{(y-z)}\,, \\
J(y)J(z) & \sim \frac{\frac{c}{3}}{(y-z)^2}\,, \\
J(y)G^{\pm}(z)& \sim \pm \frac{G^{\pm}(z)}{(y-z)}\,.
\end{align}
\end{subequations}

With the addition of $\{\widetilde{G}^{\pm}, J^{\pm \pm}\}$, the above generators obey a twisted small $\mathcal{N}=4$ SCA with central charge $c$. The remaining OPEs of this algebra are given by
\begin{subequations}
\begin{align}
J^{\pm \pm}(y) G^{\mp}(z) & \sim \mp \frac{\widetilde{G}^{\pm}(z)}{(y-z)}\,, \\
J^{\pm \pm}(y) \widetilde{G}^{\mp}(z) & \sim \pm \frac{G^{\pm}(z)}{(y-z)}\,, \\
G^{+}(y) \widetilde{G}^{+}(z) & \sim \frac{2 J^{++}(z)}{(y-z)^2} + \frac{\pd J^{++}
(z)}{(y-z)}\,, \\
\widetilde{G}^{-}(y) G^{-}(z) & \sim \frac{2 J^{--}(z)}{(y-z)^2} + \frac{\pd J^{--}(z)}{(y-z)}\,, \\
\widetilde{G}^+(y)\widetilde{G}^-(z) & \sim \frac{\frac{c}{3}}{(y-z)^3} + \frac{J(z)}{(y-z)^2} + \frac{T(z) }{(y-z)}\,, \\
T(y)J^{++}(z) & \sim \frac{ \pd J^{++}(z)}{(y-z)}\,, \\
T(y)J^{--}(z) & \sim \frac{2J^{--}(z)}{(y-z)^{2}} + \frac{ \pd J^{--}(z)}{(y-z)}\,, \\
T(y)\widetilde{G}^{+}(z) & \sim \frac{\widetilde{G}^{+}(z)}{(y-z)^2} + \frac{\pd \widetilde{G}^{+}(z)}{(y-z)}\,, \\
T(y)\widetilde{G}^{-}(z) & \sim \frac{2\widetilde{G}^{-}(z)}{(y-z)^2} + \frac{\pd \widetilde{G}^{-}(z)}{(y-z)}\,.
\end{align}
\end{subequations}
Moreover, the $\mathfrak{su}(2)_{\frac{c}{6}}$ current algebra reads
\begin{subequations}
\begin{align}
J(y) J^{\pm \pm}(z) & \sim \pm 2 \frac{J^{\pm \pm}(z)}{(y-z)} \,, \\
J^{+ +}(y) J^{--}(z) & \sim \frac{\tfrac{c}{6}}{(y-z)^2} + \frac{J(z)}{(y-z)} \,.
\end{align}
\end{subequations}

\section{Another basis for the bosonic currents}  \label{usualconventions}

\subsection{Choosing a $\rm U(1)$ direction} \label{usualconventions1}

In order to label the physical states, it is convenient to single out an $\rm U(1) \in SL(2, \mathbb{R})$ and an $\rm U(1) \in SU(2)$ direction \cite{Maldacena:2000hw}. 

We build the ${\rm SL(2,\mathbb{R})}_k$ generators in a standard basis from the currents $K_{\ab}$ by defining
\begin{align}
J_{\pm} & = - \frac{i}{2} (K_1 \pm i K_2) \,,& J_3 & = - \frac{i}{2} K_0 \,,
\end{align}
which satisfy the current algebra
\begin{subequations}
\begin{align}
J_3(y)J_3(z) & \sim - \frac{k}{2} (y-z)^{-2} \,, \\
J_3(y) J_{\pm}(z) & \sim \pm (y-z)^{-1} J_{\pm} \,, \\
J_+(y) J_-(z) & \sim k(y-z)^{-2} -2 (y-z)^{-1} J_3 \,.
\end{align}
\end{subequations}

If desired, one can do the same for the ${\rm SU(2)}_k$ part. We define the linear combinations\footnote{The choice to single out the ``five'' direction in $K_{3^\prime}=K_5$ comes because $\sg_5$ is block diagonal in our conventions, see eqs.~\eqref{6dPaulisapp}}
\begin{subequations}
\begin{align}
K_{3^\prime} & = - \frac{i}{2} K_5 \,,& K_{\pm^\prime} & = - \frac{i}{2} (K_3 \pm i K_4) \,,
\end{align}
which satisfy
\begin{align}
K_{3^\prime}(y) K_{3^\prime}(z) & \sim \frac{k}{2} (y-z)^{-2} \,, \\
K_{3^\prime}(y) K_{\pm^\prime}(z) & \sim \pm (y-z)^{-1} K_{\pm^\prime} \,, \\
K_{+^\prime}(y) K_{-^\prime}(z) & \sim k (y-z)^{-2} + 2 (y-z)^{-1} K_{3^\prime} \,.
\end{align}
\end{subequations}

\subsection{$\rm SL(2,\mathbb{R})$ and $\rm SU(2)$ quantum numbers} \label{MOvertices}

One can label the vertex operator $\mathcal{V}$ in \eqref{vertexAdS31} by the $\rm SL(2,\mathbb{R})$ quantum numbers $\{j,m\}$ and  $\rm SU(2)$ quantum numbers $\{j^\prime, m^\prime\}$. As before, we will consider zero amount of spectral flow in this section. Since all the physical degrees of freedom are contained in the superfield $V$ from eq.~\eqref{vertexAdS33}, we focus on describing its components in what follows. Moreover, the superfield $V_0 \subset \mathcal{V}$ decouples from amplitude computations presented in this work.

Accordingly, if $ \mathcal{V} \supset V $ has quantum numbers $\{j, m, j^\prime, m^\prime\}$, we write
\begin{align}
V = V\big(\vv{j},\vv{m}\big) \,,
\end{align}
where
\begin{align}
\vv{j} & = \{j, j^\prime \} \,,& \vv{m} & = \{m, m^\prime\}\,,
\end{align}
with $j= j^\prime +1$, the half-BPS condition. The vector $\vv{j}_i$ labels the $\rm SL(2,\mathbb{R}) \times SU(2)$ spin of the representation and $\vv{m}$ characterize the state in the given representation. As a consequence, under the zero-modes of the diagonal currents defined in Section \ref{usualconventions1}, we then have
\begin{align}
\vv{\nabla}_3 V & = \vv{m} V \,,& \vv{\nabla}_3 & = \{\nabla_3 , \nabla_{3^\prime}\}\,.
\end{align}

Consequently, the wavefunctions $\{ \chi_{\al 2 }, a^{\ab},\psi^{\al 2}\}$ in \eqref{vertexAdS33} can be written as
\begin{subequations}\label{halfBPSwavefunctions}
\begin{align}
\chi_{\al 2 } &=V^{\rm SL(2,\mathbb{R}) \times SU(2)}_{\vv{j} , \vv{m}_{ \al}}= V^{\rm SL(2,\mathbb{R})}_{j,m_{\al}} V^{\rm SU(2)}_{j^\prime,{m^\prime}_{\al}}\,, \\
a^{\ab} &=V^{\rm SL(2,\mathbb{R}) \times SU(2)}_{\vv{j} , \vv{m}^{\ab}}= V^{\rm SL(2,\mathbb{R})}_{j,m^{\ab}} V^{\rm SU(2)}_{j^\prime,{m^\prime}^{\ab}} \,, \\
\psi^{\al 2} &=V^{\rm SL(2,\mathbb{R}) \times SU(2)}_{\vv{j} , \vv{m}^{ \al}}= V^{\rm SL(2,\mathbb{R})}_{j,m^{\al}} V^{\rm SU(2)}_{j^\prime,{m^\prime}^{\al}}\,,
\end{align}
\end{subequations}
where $V^{\rm SL(2,\mathbb{R})}_{j_i,m_{i}}$ and $V^{\rm SU(2)}_{j^\prime_i,{m^\prime}_{i}}$ are $\rm SL(2,\mathbb{R})$ and $\rm SU(2)$ current algebra primaries, respectively. From the properties
\begin{subequations}\label{sl2su2}
\begin{align}
[\nabla_{\ab}, \ta^{\al} ]& = -{f_{\ab  \, \bt 1}}^{\al 1} \ta^{\bt}  \,,\\
[\nabla_{\ab}, (\ta \sg_{\bb} \ta) ]& = {f_{\ab \bb}}^{\cb} (\ta \sg_{\cb} \ta) \,,  \\
[\nabla_{\ab} , (\ta^3)_{\al} ] & = {f_{\ab \, \al 1}}^{\bt 1} (\ta^3)_{\bt} \,,
\end{align}
\end{subequations}
one finds that 
\begin{subequations}
\begin{align}
\vv{m}^3 & = \{m, m^{\prime} \} \,,& \vv{m}^{\pm} & = \{ m \mp 1, m^{\prime} \} \,, \\
\vv{m}^{3^\prime} & = \{m, m^\prime \} \,,& \vv{m}^{\pm^\prime} & = \{ m , m^\prime \mp 1 \} \,,
\end{align}
\end{subequations}
and $\vv{m}_{ \al}= \{m_{ \al}, m_{ \al}^\prime \}$ with
\begin{subequations}
\begin{align}
\vv{m}_{ 1} & = \{m + \tfrac{1}{2}, m^\prime - \tfrac{1}{2} \} \,,& \vv{m}_{ 2} & = \{m - \tfrac{1}{2}, m^\prime + \tfrac{1}{2} \} \,, \\
\vv{m}_{ 3} & = \{m - \tfrac{1}{2}, m^\prime - \tfrac{1}{2} \} \,,& \vv{m}_{ 4} & = \{m + \tfrac{1}{2}, m^\prime + \tfrac{1}{2} \}\,,
\end{align}
\end{subequations}
and, finally, $\vv{m}^{\al} = \{m^{\al} , {m^{\prime}}^{\al} \}$ with
\begin{subequations}
\begin{align}
\vv{m}^1 & = \{m - \tfrac{1}{2}, {m^\prime} + \tfrac{1}{2} \} \,,& \vv{m}^2 & = \{ m + \tfrac{1}{2}, {m^\prime} - \tfrac{1}{2}\} \,, \\
\vv{m}^3 & = \{ m + \tfrac{1}{2}, {m^\prime} + \tfrac{1}{2}\} \,,& \vv{m}^4 &= \{ m - \tfrac{1}{2}, {m^\prime} - \tfrac{1}{2}\}\,.
\end{align}
\end{subequations}

\section{Gauge invariance in the hybrid description} \label{gaugeinvhybV}

Let us analyze the consistency of the hybrid vertices \eqref{3pointterms} appearing in the three-point amplitude for the states in the NS-sector. We take $V= \frac{i}{2} (\ta \sg_{\ab} \ta) a^{\ab} $, so that eqs.~\eqref{3pointterms} become
\begin{subequations}\label{hybridwavefs}
\begin{align}
\mathcal{V}& = e^{\rho + i \sg} \frac{i}{2} (\ta \sg_{\ab} \ta) a^{\ab} \,, \\
\widetilde{G}^+_0 \mathcal{V} & = e^{2 \rho + i \sg} J^{++}_C \frac{i}{2} (\ta \sg_{\ab} \ta) a^{\ab} \,, \\
G^+_0 \mathcal{V} & = -\frac{1}{2k} \frac{1}{\sqrt{2}} e^{i \sg} \bigg( K_{\ab} a^{\ab} + i S_{\al 1} (\sg^{\ab \bb} \ta)^{\al} D_{\ab} a_{\bb} + i \sqrt{2} S_{\al 1} (\dth \sg_{\ab} \ta)^{\al} a^{\ab} \bigg) \,. 
\end{align}
\end{subequations}
Up to a constant, the integrated vertex operator is then given by
\begin{align}\label{intvertexhyb}
\int G^-_{-1}G^+_0 \mathcal{V} & =  \int \big(K_{\ab} a^{\ab} + i S_{\al 1} (\sg^{\ab \bb} \ta)^{\al} D_{\ab} a_{\bb} + i \sqrt{2} S_{\al 1} (\dth \sg_{\ab} \ta)^{\al} a^{\ab} 
\big)\,.
\end{align}

We now check that the integrand of \eqref{intvertexhyb} is gauge invariant up to a total derivative. From eq.~\eqref{gaugetransfAdS3}, one finds that the gauge transformation for $a_{\ab}$ is
\begin{align}\label{gaugetransfAdS3NSsec}
\dt a_{\ab} & = \nabla_{\ab} \la \,,
\end{align}
for some $\la = \la (g)$ where $g \in {\rm PSU(1,1|2)}$ and $\nabla_{\al 1} \la =0$. Therefore, under \eqref{gaugetransfAdS3NSsec}, the integrated vertex operator \eqref{intvertexhyb} transforms as
\begin{align}
& K^{\ab} \nabla_{\ab} \la - S_{\al 1} \ta^{\bt} {f_{\ab \, \bt 1}}^{\al 1} \nabla^{\ab} \la \nn \\
& \qquad = K^{\ab} \nabla_{\ab} \la + S^{\al j} \nabla_{\al j} \la \nn \\
& \qquad = \pd \la\,,
\end{align}
as we wanted to show. In arriving to the second line above we used that
\begin{align}
\nabla_{\al 1} \la  = 0 \qquad \Rightarrow \qquad S^{\al 2} \nabla_{\al 2} \la  = - S_{\al 1} \ta^{\bt} {f_{\ab \, \bt 1}}^{\al 1} \nabla^{\ab} \la\,.
\end{align}

As a result, we conclude that the integrated vertex operator \eqref{intvertexhyb} is gauge invariant up to a total derivative in the group manifold of $\rm PSU(1,1|2)$.
 

\bibliography{bibliomain}
\bibliographystyle{JHEP.bst}

\end{document}